\documentclass[letterpaper,twocolumn,10pt]{article}
\usepackage{usenix-2020-09}
\usepackage{epsfig,endnotes, multirow,}
\usepackage{cite}
\usepackage{amsmath,amssymb,amsfonts}
\usepackage{algorithmic}
\usepackage{graphicx}
\usepackage{textcomp}
\usepackage{xcolor}
\usepackage{subcaption}
\usepackage{filecontents}
\usepackage{epsfig,endnotes, multirow}
\usepackage{array}
\usepackage{subcaption}
\usepackage{color}
\usepackage{url}
\usepackage{hyperref}
\hypersetup{colorlinks=true,breaklinks=true}
\usepackage{breakurl}

\usepackage{rotating}

\microtypecontext{spacing=nonfrench}
\newcommand{\eg}{\textit{e}.\textit{g}.,~}
\newcommand{\ie}{\textit{i}.\textit{e}.,~}

\definecolor{darkgreen}{RGB}{0,120,0}
\definecolor{darkblue}{RGB}{0,0,200}
\definecolor{airforceblue}{rgb}{0.36, 0.54, 0.66}
\definecolor{bleudefrance}{rgb}{0.19, 0.55, 0.91}
\definecolor{cerulean}{rgb}{0.0, 0.48, 0.65}

\definecolor{bluegray}{rgb}{0.4, 0.6, 0.8}

\newcommand{\revision}[1]{\textcolor{black}{#1}}
\newcolumntype{L}[1]{>{\raggedright\let\newline\\\arraybackslash\hspace{0pt}}m{#1}}
\newcolumntype{C}[1]{>{\centering\let\newline\\\arraybackslash\hspace{0pt}}m{#1}}
\newcolumntype{R}[1]{>{\raggedleft\let\newline\\\arraybackslash\hspace{0pt}}m{#1}}

\newcommand{\squishlist}{
   \begin{list}{$\bullet$}
    { \setlength{\itemsep}{0pt}      \setlength{\parsep}{3pt}
      \setlength{\topsep}{3pt}       \setlength{\partopsep}{0pt}
      \setlength{\leftmargin}{1.0em} \setlength{\labelwidth}{1em}
      \setlength{\labelsep}{0.5em} } }
\newcommand{\squishend}{
    \end{list}  }

\begin{document}

\date{}

\title{\Large \bf That Doesn't Go There: Attacks on Shared State in \\ Multi-User Augmented Reality Applications}
\author{
{\rm Carter Slocum\thanks{Equal contributors.}~~$^1$, Yicheng Zhang{\color{green!80!black}$^*$}$^1$, Erfan Shayegani$^1$, Pedram Zaree$^1$, Nael Abu-Ghazaleh$^1$, Jiasi Chen$^2$}\\
$^1$University of California, Riverside\\
$^2$University of Michigan
}

\maketitle

\pagestyle{empty} 


\begin{abstract}

Augmented Reality (AR)
can enable shared virtual experiences between multiple users. In order to do so, it is crucial for multi-user AR applications to establish a consensus on the ``shared state'' of the virtual world and its augmentations through which users interact. 
Current methods to create and access shared state collect sensor data from devices (\eg camera images), process them, and integrate them into the shared state.
However, this process introduces new vulnerabilities and opportunities for attacks.
Maliciously \emph{writing} false data to ``poison'' the shared state is a major concern for the security of the downstream victims that depend on it. 
Another type of vulnerability arises when \emph{reading} the shared state: by providing false inputs, an attacker can view hologram augmentations at locations they are not allowed to access. 
In this work, we demonstrate a series of novel attacks on multiple AR frameworks with shared states, focusing on three publicly accessible frameworks.
We show that these frameworks, while using different underlying implementations, scopes, and mechanisms to read from and write to the shared state, have shared vulnerability to a unified threat model. Our evaluations of these state-of-the-art AR frameworks demonstrate reliable attacks both on updating and accessing the shared state across different systems.  
To defend against such threats, we discuss a number of potential mitigation strategies that can help enhance the security of multi-user AR applications \revision{and implement an initial prototype}.
\end{abstract}


\section{Introduction}
\label{sec:intro}

AR technologies have enabled a large variety of applications that use real-world data to create environments enriched with overlaid virtual holograms. These virtual holograms can take many forms, from face filters to virtual characters, and they are typically placed relative to some point in the real world, such as a table, face, or recognizable landmark. Although AR has been around for several decades~\cite{CaudellMizell}, the recent ubiquity of mobile devices and the availability of commercial AR headsets have made it possible for AR applications to reach the mass market~\cite{nytPokemonGo}. 
Recent AR applications even allow multiple users to interact with the same AR holograms. 
For example, in 2019, Pok\'emon Go enabled users to view the same virtual creatures at the same time in some shared space using a ``Buddy Adventure'' system~\cite{BuddyAdventure}. 
In order for these multi-user interactions to take place, some information about the state of the real world (\eg nearby flat planes, landmarks, and virtual objects) must be sensed, processed, and shared between users to provide a common frame of reference.
We call this information, together with the hologram information, as the ``shared state'' of the AR application.
Several multi-user AR systems with cloud-based AR shared states exist and are in use, including those by Google~\cite{cloud_anchor_intro} and Meta~\cite{Mapillary}.
Thus, a natural question emerges after the rise of such systems: \textbf{What security threats can exist for AR frameworks involving this shared state?}

In this work, the attacks that we focus on relate to one of the fundamental problems in AR: how to place a hologram accurately in the real world and have it persist over time, space, and across users.
Successful manipulations of hologram locations could have serious impacts on both owners and users of the system.  As the number of users and businesses relying on AR continues to increase, the incentives for attackers to manipulate the shared state to their advantage also increase.  For example, suppose a construction company is using AR to place and visualize markings in the environment. \revision{Construction workers wearing AR glasses might visit the site and} visualize where a water pipe should be built or where to dig a hole.
A vulnerable AR construction application could cause confusion, destruction of property, or danger to workers if an attacker's efforts result in a \revision{construction worker viewing virtual demolition markers in unsafe real-world areas and bulldozing those areas.}

Our first goal is to identify the threat models that affect the shared state.  At a high level, interactions between users and shared state in AR can be thought of as \emph{read} and \emph{write} operations, which are provided by AR frameworks through an API.
One can write a new virtual hologram to the shared state and read others' virtual holograms in order to render them on the device.
Direct manipulation of the shared state is impossible since it's typically stored on cloud/edge servers controlled by the AR service and is well-secured, requiring physical or software exploitation not unique to AR. Instead, we investigate exploits that allow attackers to remotely manipulate the shared state using only the basic read/write API calls available to users.
Calls to the API typically involve associated location data consisting of one or more of the following: Global Positioning System (GPS) coordinates, camera images, and/or Inertial Measurement Unit (IMU) sensor data.  This information allows the attacker to map the read or write operation to a location within the AR space.

We seek to understand these threats and develop end-to-end attacks on commercial systems, in order to demonstrate how they work and inform designers and develop mitigations. 
We develop a range of attacks targeting write or read operations on the Google's Cloud Anchor API~\cite{cloud_anchor_intro}, Google's ARCore Geospatial API~\cite{Geospatial}, and Meta's Mapillary system~\cite{Mapillary}.
\revision{We focus on these because they are major commercial players, and their multi-user APIs work on many common AR devices.}
We show that malicious reads and writes are possible on the three systems despite the substantial differences in how they perform these operations.  We are able to write information to different, potentially inaccessible, locations on the map, as well as falsify our own location to access information at potentially private or inaccessible locations.
These end-to-end attacks rely on known image and GPS spoofing methods; our contribution lies in combining these methods with multi-user AR, which brings unique challenges in that the interactions between the real world, virtual world, and downstream victims must be accounted for in a successful attack.

In summary, we make the following contributions:
\squishlist
\item We create a taxonomy of existing commercial AR frameworks with shared state
and
identify their common vulnerabilities regarding the read and write operations. 
We form a unified threat model that covers these current and prospective AR applications.
    \item \revision{We demonstrate attacks on shared state in the novel AR domain on three frameworks using real devices (smartphones), and quantify their success.} 
To the best of our knowledge, \revision{this is the first documented demonstration of such attacks on AR using these frameworks.} 
\item We repeat the attacks of these three scenarios in various environments (\eg different locations, lighting, clutter) to demonstrate the attack's robustness.
\item \revision{We propose and evaluate a defense strategy that uses multi-modal sensors (depth and visible light cameras) available on some AR devices. The source code and dataset are available online.}\footnote{ \url{https://sites.google.com/view/multi-ar-defense/}}
\squishend


\paragraph{Disclosures and ethics.} We disclosed our findings to Google and Meta.
\revision{We performed all experiments either in sandboxes or in our own local sessions, so no external public users were affected by our experiments.}

%


\section{Background}
\label{sec:Background}


In this section, we first introduce the background of shared state in AR (Section~\ref{sec:sharestatear}).
We then describe the current landscape of shared state in commercial AR systems (Section \ref{sec:taxonomy}).
Finally, we define the general threat model (Section~\ref{sec:threat}).

\subsection{Shared State in Augmented Reality}
\label{sec:sharestatear}


\begin{figure}
    \centering
    \includegraphics[width=0.45\textwidth]{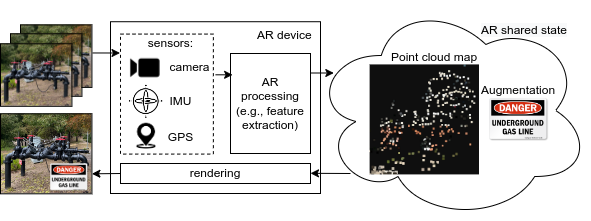}
    \vspace{-0.1in}
    \caption{AR processing pipeline. An AR device senses the environment, processes the sensed data, and uploads information to the shared state. The shared state returns an augmentation overlaid onto the user's display. 
    }
    \label{fig:device_pipeline}
\end{figure}


To facilitate interactions between multiple users in AR, a mutually agreed-upon model of the reality to augment, and the augmentations within it, is needed between users~\cite{MultiConsistent, MultiHandheld}.
Ideally, this model should be consistent across devices and thus is typically stored in the cloud, providing a central access point.
In such a model, multiple users interact with the shared augmentations (\eg other participants in a remote meeting app). 
They also fuse spatial information about the real environment using the collected sensor data.
We call this shared model of reality the \textit{shared state}.

The shared state commonly contains a ``map'' of 3D points (an example is shown on the right side of Fig.~\ref{fig:device_pipeline}).
The points in this map are features extracted from images (\eg \cite{ng2003sift,rublee2011orb}).
Each feature contains an estimate of its 3D position and a descriptor of its visual neighborhood for use in finding and correlating the same feature in other images.
To give augmentations the appearance of blending in with the real world, they are described by their 3D coordinates.
Thus, the AR shared state consists of the map of visual features combined with the augmentations placed on the map.
Fig.~\ref{fig:device_pipeline} shows the processing pipeline of an AR device accessing the shared state in the cloud, including communicating with the shared state to receive augmentations and render them onto the display.

\paragraph{Communication with the shared state.} 
For a user to view or place shared holograms/augmentations, communication with the shared state is needed.
Abstractly, we can think of viewing or placing the shared holograms as read and write operations against the shared state, respectively, using key-value pairs.
The \emph{key} is some piece of information relating to the user's physical location, which a user provides (details later), and the \emph{value} is the associated hologram's coordinates (and optionally its visual appearance).
The cloud processes these key-value pairs and updates (or retrieves information from) the shared state accordingly.
There are two operations for users to communicate with the shared state: \emph{read} and \emph{write}, as follows.

\begin{itemize}
    \item \textbf{Read:} A user may read the shared state to determine where she is on the map and render the appropriate holograms. For instance, a user may go to a park where virtual art is displayed and upload an image \emph{key} of the park to the cloud, captured by the phone's camera, and receive back the \emph{value} of the hologram's coordinates, and then render the virtual art on her display. 
    \item \textbf{Write:} Users may write holograms at specific locations in the map in the shared state. 
    For instance, a user may place their own virtual art for other users to view by uploading a \emph{key} consisting of a short image sequence near the art and the associated GPS coordinates alongside a \emph{value} of the virtual art's coordinates.
\end{itemize}

Keys consist of information used to identify locations within the shared state.  Keys are usually derived from three main types of sensors commonly used in AR applications: GPS, camera, and IMU. GPS data provides information about the user's geographical location and typically consists of latitude, longitude, altitude, and time. Camera data in AR applications can take the form of video or a sequence of timestamped images. IMU data refers to the measurements collected by sensors such as accelerometers, gyroscopes, and magnetometers. This data provides information about the device's orientation, acceleration, and rotation. The IMU may not be strictly necessary for these applications to work but is often included to assist in speed and accuracy~\cite{VisInertSLAM}. 

\subsection{AR Shared State Taxonomy}
\label{sec:taxonomy}

\begin{table}[t]
\small
\centering
\resizebox{0.46\textwidth}{!}{%
\begin{tabular}{C{1cm}|L{3.2cm}||L{3.2cm}|}
\cline{2-3}
\textbf{}                                              & \textbf{Non-curated}     & \textbf{Curated}               \\ \hline \hline
\multicolumn{1}{|l|}{\multirow{3}{*}{\textbf{Local}}}  & \textbf{Scenario A: Cloud Anchor } &  \textbf{Commercial scenario not found. }\\
\multicolumn{1}{|l|}{}                                 & \textit{Keys}: camera, IMU        & \textit{Keys}: camera, IMU              \\
\multicolumn{1}{|l|}{}                                 & \textit{Attacks}: read, write     & \textit{Attacks}: read                  \\ \hline
\multicolumn{1}{|l|}{\multirow{3}{*}{\textbf{Global}}} & \textbf{Scenario C: Mapillary               }    &  \textbf{Scenario B: Geospatial Anchor } \\
\multicolumn{1}{|l|}{}                                 & \textit{Keys}: camera, IMU, GPS   & \textit{Keys}: camera, IMU, GPS         \\
\multicolumn{1}{|l|}{}                                 & \textit{Attacks}: write           & \textit{Attacks}: read                  \\ \hline
\end{tabular}%
}
\caption{Taxonomy of AR shared states. 
}
\label{tab:attack_taxonomy}
\end{table}


We studied the current landscape of multi-user AR and found three major examples of shared state: Cloud Anchor~\cite{cloud_anchor_intro}, Geospatial Anchor~\cite{Geospatial}, and Mapillary~\cite{Mapillary}, which we primarily focus on in this work.
Cloud Anchor and Geospatial Anchor are part of Google ARCore, which is Google's AR Software Development Kit (SDK) for Android devices.
Mapillary is a crowd-sourced mapping service acquired by Meta in 2020.
These frameworks abstract away low-level details so that developers can more easily build AR applications on top, so vulnerabilities in the underlying frameworks will affect many AR applications.
The design of these frameworks can be dissected along two dimensions: global/local and curated/non-curated, as summarized in Table~\ref{tab:attack_taxonomy}.
Next, we describe each of these dimensions.

\paragraph{Global vs. local shared state.}
AR applications can run in local or global geographic areas; for example, a treasure hunt may take place locally within a building, while Pok\'emon Go takes place globally.
Consequently, they can have larger or smaller maps in their shared state,  which we categorize as a global or local shared state.
AR frameworks with the global shared state tend to utilize GPS coordinates plus camera images as the key to writing
into the shared state. 
Specifically, each writer uploads local images tagged with GPS coordinates to the shared state, where the cloud merges all data 
to create a global shared state.
Users seeking to read from the shared state may use a combination of GPS, camera, and, optionally, IMU data as a key into the database.
Global shared states tend to be persistent without a clear expiry time, typically persisting for years. 

AR frameworks with local shared states are typically smaller in geographic scope 
and lack global positioning (GPS).
The key typically consists of just camera images and optional IMU data, without GPS. 
Local shared states tend to be ephemeral in that they have a configurable lifetime, typically of less than one year~\cite{cloud_anchor_intro}.

\paragraph{Curated vs. non-curated shared state.}
The maps contained in the shared state can be either curated or non-curated.
Curated maps are constructed by ``high trust'' users or ``curators''.
These curators have elevated write permissions to the shared state and usually have the incentive to avoid malicious behavior. 
Most commonly, these curators are paid employees, contract workers, or trusted research groups.  
An example is the Street View Car~\cite{street-view-car}, where company employees drive a car around and capture camera images to upload to the cloud, which processes them and inserts them into the shared state's map.
Non-curators can still read the curated shared state but cannot otherwise manipulate it.

AR frameworks with non-curated (\ie crowd-sourced) shared states allow all users to read and write to the map in the shared state. 
These users are low trust but come with the advantage of increased numbers, allowing rapid construction and updating of the shared state compared to curators.
An example is Mapillary's crowd-sourced street mapping model, where public users can upload camera images to the cloud, which processes them and inserts them into the map.

The write permissions for the shared state maps and the shared state holograms may be separate.
For our purposes, a curator has permission to write both shared state map and hologram data to the shared state, while a non-curator can only read map data from the shared state but may be able to both read and write holograms.
In the future, applications with more granular permissions may become more common~\cite{claramunt2023spacemediator}.


\subsection{Threat Model}
\label{sec:threat}
\begin{figure}[t]
    \centering
    \begin{subfigure}[b]{0.42\textwidth}
         \centering
         \includegraphics[width=\textwidth]{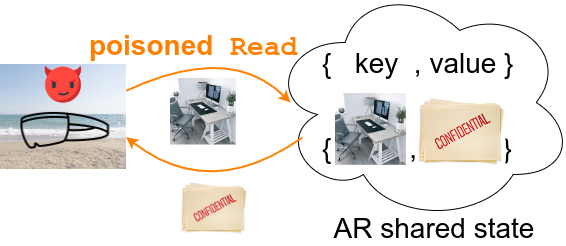}
         \caption{Read attack.}
         \label{fig:read_attack}
     \end{subfigure}
     \begin{subfigure}[b]{0.42\textwidth}
         \centering
         \includegraphics[width=\textwidth]{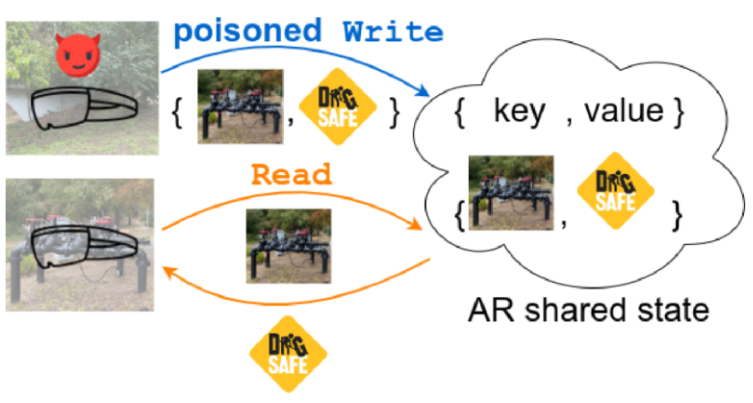}
         
         \caption{Write attack.}
         \label{fig:write_attack}
     \end{subfigure}
    \caption{Attacks on AR shared state. Read attack: A \revision{private} hologram is read outside the area it was written to (beach instead of office). Write attack: A hologram is written to an area where the attacker is not present (pipes instead of field). 
    }
    \label{fig:threatoverview}
\end{figure}

We assume an attacker engages in AR experiences with shared states using an unmodified AR application. The attacker 
only possesses the same read/write permissions as normal users. The primary objective of the attacker is to compromise the integrity or confidentiality of the multi-AR shared state. We identify two classes of attacks in this context (see Fig.~\ref{fig:threatoverview}): \textit{(a) read attack} and \textit{(b) write attack}.

\paragraph{Read attack.} Such an attack focuses on extracting sensitive information stored within the shared state created by other users. 
For example, suppose a victim user has created a \revision{hologram of a whiteboard and written sensitive company secrets onto it. The whiteboard} is uploaded to the shared state and is only supposed to be viewable from the private office. Thus, in Fig.~\ref{fig:read_attack}, the shared state contains the \{key=office image, value=confidential whiteboard document\} entry. The objective of the attacker is to retrieve and access this private document, thereby breaching confidentiality, by providing a forged \{key=office image\} 
to retrieve the associated value (the private hologram). \revision{The attacker benefits from retrieving confidential information, which is a serious concern in multi-user AR~\cite{ruth2019secure}.
It is assumed that the attacker can gain temporary physical access to the office in order to obtain images to use later during the attack, or is able to find publicly available images of the office.}

\paragraph{Write attack.} The attacker seeks to manipulate the shared state in order to deceive subsequent victim AR users. Specifically, the attacker creates and uploads manipulated images or falsified sensor readings as keys in the shared state, and uses them to add to the shared state at that location without being there. 
\revision{It is assumed that the attacker possesses images and GPS coordinates of the target location needed for manipulation.}
Thus, in Fig.~\ref{fig:write_attack}, the shared state contains the \{key=pipe image, value=``dig safe'' sign\} entry created by the attacker. 
Subsequently, when victims attempt to read from the shared state, they may encounter misleading or false information, leading to inaccurate perceptions or actions within the AR environment. For example, in Fig.~\ref{fig:write_attack}, the victim uses a legitimate \{key=pipe image\} and retrieves a hologram telling her it is safe to dig there. 
\revision{The attacker benefits by causing disruption of legitimate AR use cases.}
Moreover, with companies now making efforts to combine their maps (\eg Overture Maps Foundation~\cite{Overture} includes Amazon, Meta, and Microsoft as contributors), poisoned writes to one shared state could potentially propagate to other shared states.

\paragraph{Key issues.} The fundamental issue with the shared state that enables these attacks is that the ingest pipelines of these AR frameworks accept most keys as inputs. They do not have a way of verifying that users are uploading legitimate information consistent with the key they provide. Furthermore, even if the attacker fails to generate perfect keys identical to legitimate inputs, the shared state still accepts them because it attributes their imperfections to noise.
We speculate that these weaknesses are due to the nascent nature of multi-user AR frameworks; because AR frameworks want to encourage user participation, they favor functionality and lowering barriers to participation over security. 
\revision{
The collaborative nature of these applications necessitates opening a shared state for read and possibly write access among large groups of users that are not necessarily mutually trusting.}



\paragraph{Attacker’s goal in each scenario.} 
As various multi-AR platforms rely on different combinations of sensor inputs to generate these keys, our investigation focuses on three attack scenarios outlined in Table~\ref{tab:attack_taxonomy}. 

In Scenario A, the attacker's goal is to perform both read and write attacks on the shared state.
It aims to read or write holograms to locations where they are not physically present. By doing so, it deceives other users by providing false or manipulated information. 
Since AR applications in such a scenario run in local areas only, the attacker only needs camera and IMU data as keys to read or write from the shared state, and not any global information (GPS), making this attack easier to realize.
\revision{
We assume that the attacker can participate as a regular user in the AR session, which can be protected by API credentials and a room code.
API credentials are commonly hard coded into the app, so the attacker does not need to learn it, and the room code is an integer that by default starts at 1 and increments every session, making it feasible for the attacker to find through brute force~\cite{cloud_anchor_intro}.}

In Scenario B, the attacker's goal is to perform a read attack only. It attempts to read a hologram from a location where the hologram does not exist, effectively lying about her location and reaping the benefits. 
In addition to the camera and IMU data needed as keys in Scenario A, the global nature of this scenario requires the attacker to understand the global position of the hologram she wishes to read, necessitating GPS data in the key.
We do not investigate write attacks in Scenario B due to the curated nature of the shared state.
In other words, since the threat model assumes the attacker is an ordinary user, only read attacks can be performed on a curated shared state with the appropriate key.
These keys are used by all users freely with no need for special permissions.
\revision{We assume that the attacker has the API credentials hardcoded in the app, as in Scenario A, but it does not require a local room code because the shared state is global. } 

Finally, the Scenario C attacker writes holograms to false locations. This would allow an attacker to manipulate holograms that other users view, potentially leading to sabotage and safety issues.
The attacker's writes are uniquely enabled by the non-curated nature of the shared state in this scenario.
Again, special attention must be paid to the global positions of the holograms and map data for successful attacks due to this scenario's global scale.
We do not investigate read attacks in Scenario C because this API does not yet exist in the commercial framework we studied.
\revision{The attacker does not require any special credentials or room codes because the shared state is global and crowd-sourced.}

We did not find any current examples of an AR framework that provides a local and curated shared state (upper right box in Table~\ref{tab:attack_taxonomy}). 
We speculate that such a shared state could be created by a local administrator who curates the map and holograms in small areas, such as a university campus. 
Related frameworks also exist in the research domain~\cite{chen2018marvel}.





\section{Scenario A: Local, Non-Curated Shared State}
\label{sec:cloudanchor}
In this section, we focus on attacks on AR frameworks with local and non-curated shared states.
In particular, we focus on the Cloud Anchor API~\cite{cloud_anchor_intro}, which allows multiple users to share experiences within a single app. It is the underlying mechanism enabling multi-user AR apps on Android devices.
Using an app that integrates this API, a user (User A) can write a hologram to a specific location within their environment, such as the surface of a desk. 
Another user (User B), who has access credentials to the app \revision{(see \S\ref{sec:threat})}, can then read the hologram from the shared state and view and interact with it in the same physical space. 
We identified an attack vector related to this multi-user functionality, described in the following subsection.
\revision{The experiments were performed with our own test devices in a private local session only, so no external users were affected by our experiments.}


\subsection{Methodology}
\label{sec:cloudanchor-method}



The normal process of writing a hologram to the shared state involves User A pointing her AR device at the desired location of the hologram (\eg a desk) and moving around it to capture the required keys (camera images and IMU readings), which are uploaded to the shared state along with the hologram.
If User B wants to read the hologram uploaded by the previous user, she points her device at the same location, captures a key, and sends it to the shared state. 
If the key matches an entry in the shared state, the corresponding value (hologram uploaded by User A) is retrieved from the cloud, and User B can view it.
If there is no matching key, the API rejects User B's read request.

Normally, successful reads and writes require the user to be physically present in the environment where the hologram was placed in order to generate the correct corresponding key. However, our attack disrupts this workflow and demonstrates that attackers can remotely launch read and write attacks. Specifically, we show that the attacker can perform these actions using only an image of the environment (\eg printed on a photograph or displayed on a computer monitor). By pointing the camera at the image, the attacker deceives the API into believing that it is physically located in the environment, even though it is not actually present. Next, we describe three sub-types of this general attack\revision{: read, write, and triggered write. We focused on read and write because they are fundamental primitives, and triggered write is an advanced version with more targeted attack timing.}
Further technical details on the methodology of this attack and others are provided in Appendix~\ref{app:experiment_procedure}.

\begin{figure}
  \begin{center}
    \includegraphics[width=0.42\textwidth]{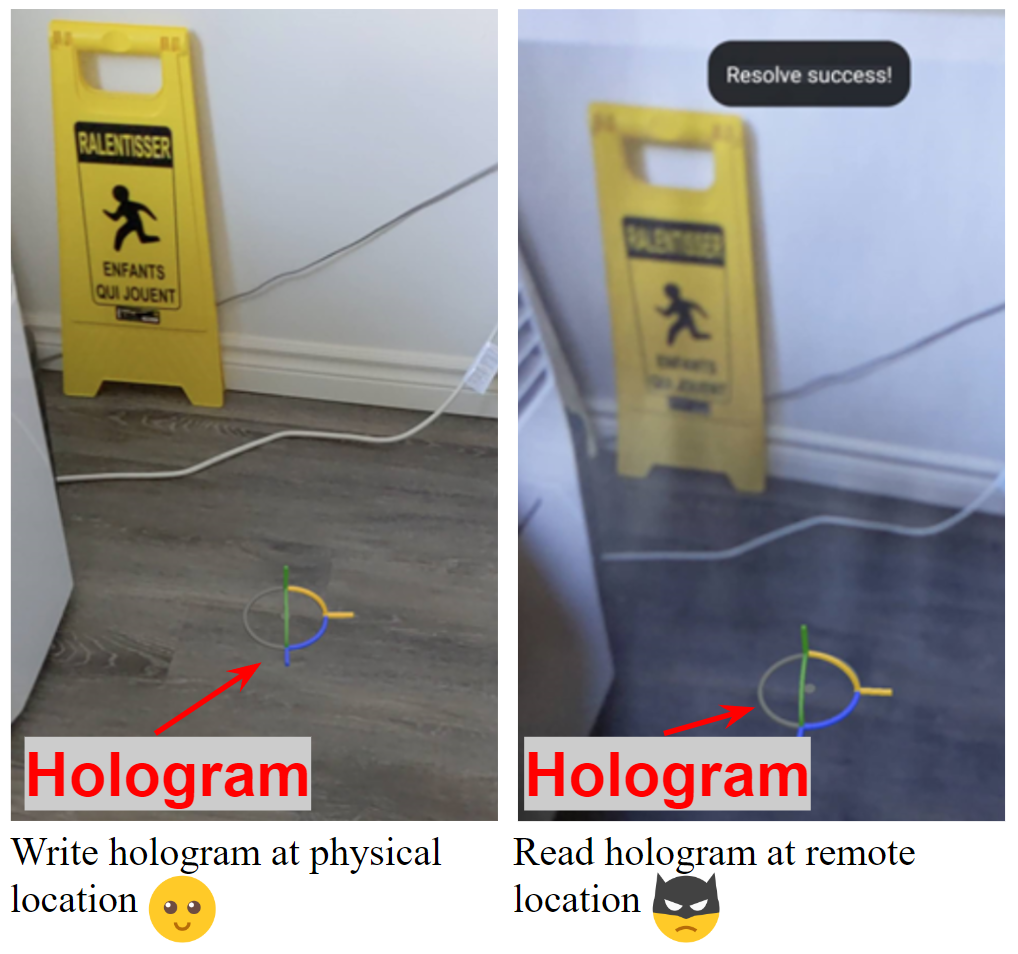}
  \end{center}
  \vspace{-0.2in}
  \caption{Remote read attack in Scenario A. \emph{Left:} A victim places a hologram in front of a yellow sign. \emph{Right:} An attacker is able to view the hologram from a photograph without being physically near the yellow sign. }
  \label{fig:cloud-read}
\end{figure}



\paragraph{Remote read attack.} In this attack, an attacker can remotely read a hologram, different from where a victim originally placed the hologram. 
\revision{The attacker benefits because it can read private notes, passwords, or even sound files that belong to the victim.}
We assume the attacker has the pre-knowledge of the victim's physical location. For instance, the attacker may have a chance to view an image of the victim's office.
The attacker's methodology is simple yet effective: it prints physical photographs or displays virtual images of the location where a hologram is placed and moves the AR device around to view the photograph/display from slightly different angles.
This generates the necessary key (camera images and IMU readings) to retrieve the hologram from the shared state.
Both the camera images and IMU readings (orientation of the device) must reasonably match the key previously stored in the shared state by the victim.
The attacker's read request may fail if the camera image differs significantly (\eg zoomed out) from where the victim originally wrote the holograms or if the IMU readings differ (\eg the victim wrote the hologram while the device was in landscape mode but the attacker tried to read the holograms from portrait mode).
Fig.~\ref{fig:cloud-read} shows an example of such an attack (``Resolve success!'' means attack succeeds). 
The hologram (a colorful 3D axis) is initially placed in front of the yellow sign by a victim.
Later, an attacker with a photograph of the yellow sign can view the hologram, despite being at a different location and nowhere near the yellow sign.

\begin{figure}
  \begin{center}
    \includegraphics[width=0.42\textwidth]{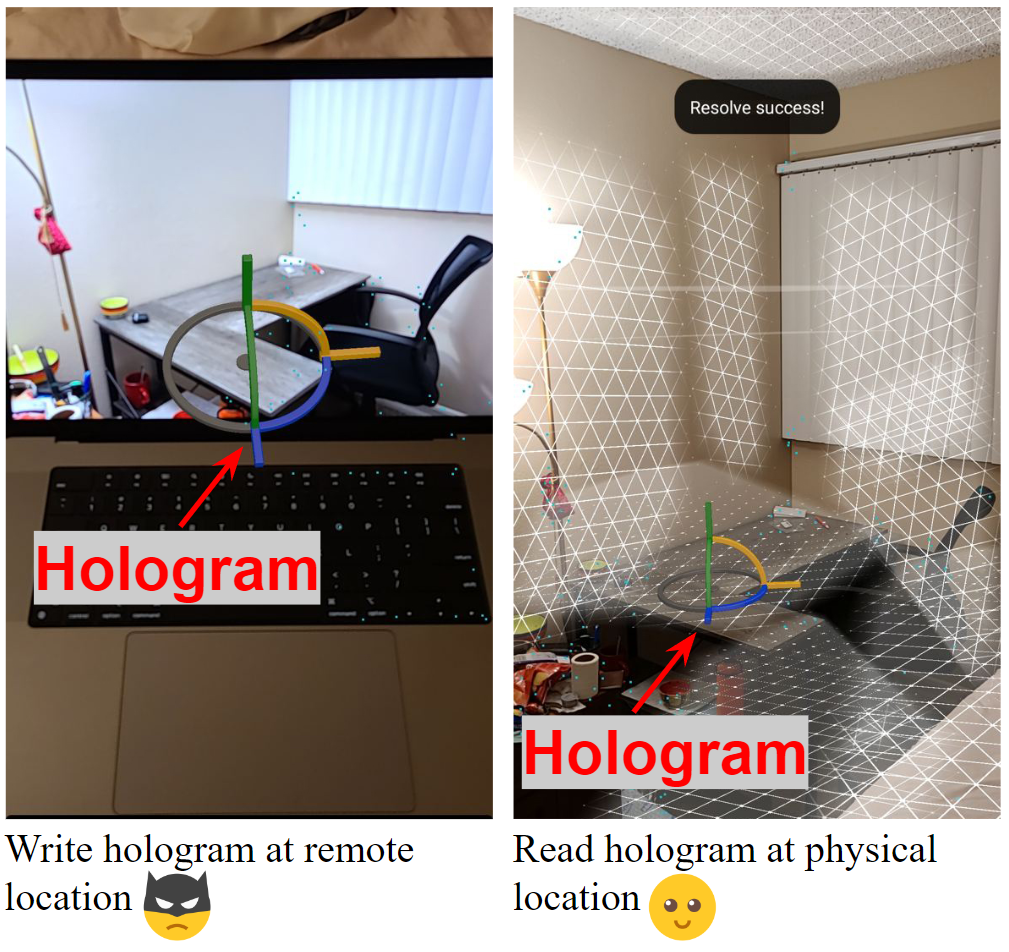}
  \end{center}
  \vspace{-0.2in}
  \caption{Remote write attack in scenario A.
  \emph{Left:} An attacker is able to write a hologram at a real-world location (a desk) without being physically present. \emph{Right:} A victim views the unexpected hologram on the desk.
  }
  \label{fig:cloud-write}
\end{figure}

\paragraph{Remote write attack.}  In this type of attack, an attacker can write AR holograms in places where it is not authorized to access or contribute, such as holy sites, museums, private spaces, kindergartens, and more. This situation becomes even more concerning if the written AR holograms contain inappropriate material, such as racist, extremist, pornographic, or disturbing content, \revision{causing psychological harm to victims}.
The attacker's methodology is similar to the remote read attack, with the additional step that after viewing the photograph/display, the attacker also indicates (by interacting with the AR device) where within the photograph/display the hologram should be placed.
However, a significant challenge was that the key the attacker needs to generate a write request is more detailed than that needed for a read request; more camera images of the scene need to be captured from different angles to generate a successful write request (while looking at a single image on the display). 
We successfully tackled this challenge by carefully maneuvering the camera and prioritizing the capture of the image displayed on the monitor while minimizing the inclusion of the surrounding environment.

In Fig.~\ref{fig:cloud-write}, we show an example of this attack.
The attacker displays an image of a desk on a computer monitor and places (writes) the hologram onto the desk in the shared state. When the victim later visits the location and views the desk through her AR device, it retrieves the hologram maliciously written by the attacker.
Note that the key of the attacker's write request did not have to exactly match the key of the victim's read request, as illustrated by the differences in the camera images of the attacker and victim (compare the left and right sides of Fig.~\ref{fig:cloud-write}); for example, the attacker had extra features such as the keyboard and the monitor's border in view.
The attack was still successful because the shared state matches keys that are not perfectly identical to allow for legitimate scenarios, such as two users viewing the same scene from slightly different angles.

\begin{figure}
  \begin{center}
    \includegraphics[width=0.45\textwidth]{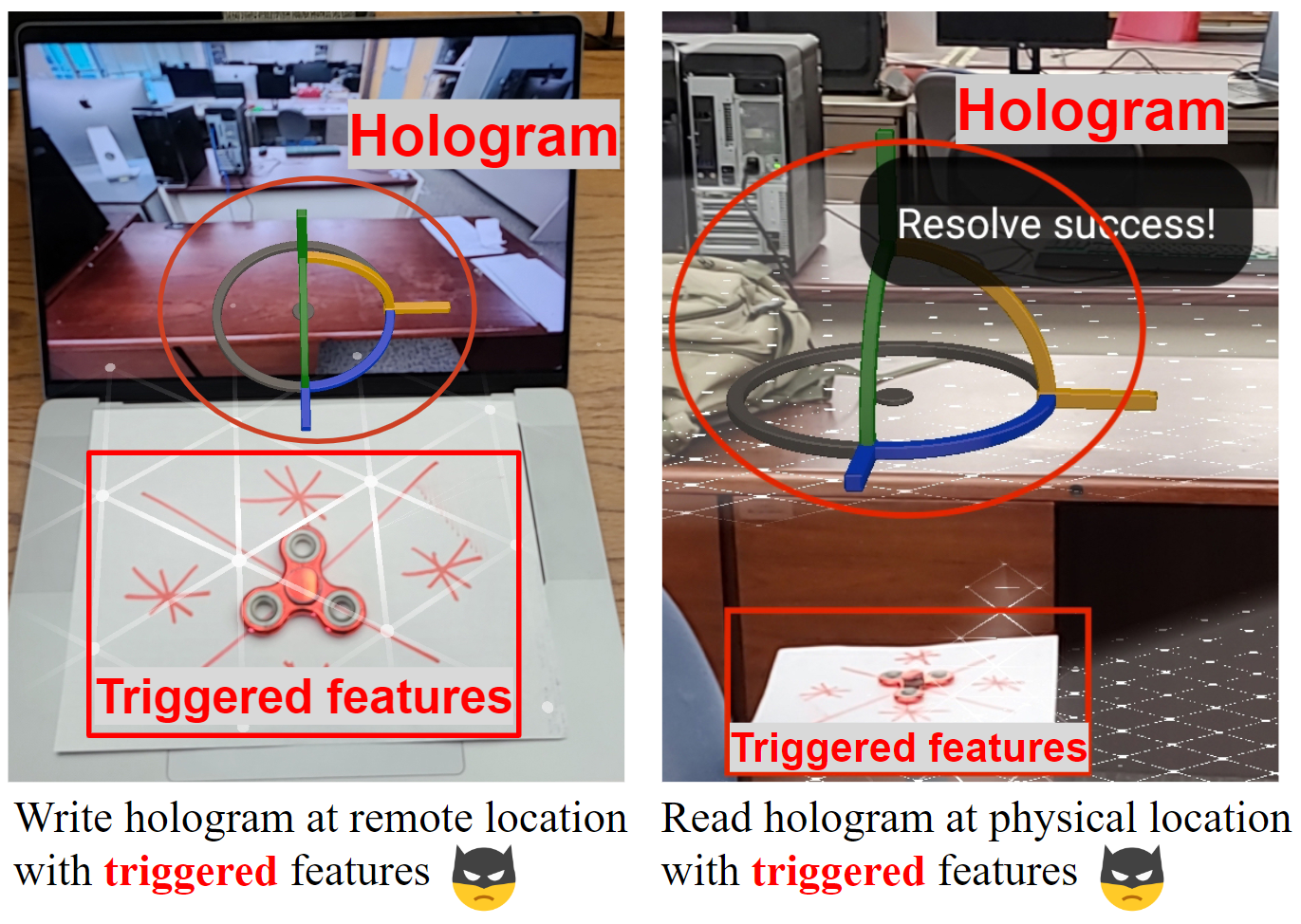}
  \end{center}
  \vspace{-0.2in}
  \caption{Triggered remote write attack in scenario A. 
  \emph{Left:} An attacker employs triggered features to remotely write a hologram at a real-world location without being physically present.
  \emph{Right:} A victim encounters an unexpected hologram on their desk, triggered by features injected by the attacker.}
  \label{fig:trigger-attack}
\end{figure}

\paragraph{Triggered remote write attack.} This attack can be treated as an advanced 
type of write attack, but it is more stealthy.
We assume that an attacker not only has the ability to execute a successful remote write attack and poison the shared state, but it also has the ability to manipulate the victim's environment with pre-determined triggered features. This allows the attacker to exert control over the timing and extent of the attack, targeting specific individuals or groups.
 For instance, consider a scenario where a TV is present in the environment.
 Suppose the attacker could strategically turn on the TV and display the trigger on the screen when a specific person enters. This greatly increases the probability of the victim's successful reading and display of the attacker's hologram, leading to potentially severe consequences as desired by the attacker. 

Figure~\ref{fig:trigger-attack} illustrates this, where the attacker initially writes a hologram remotely with the triggered features (left side of Fig.~\ref{fig:trigger-attack}). Subsequently, if the attacker places the same triggered features at the victim's physical location (right side of Fig.~\ref{fig:trigger-attack}), the victim will read the hologram placed by the attacker from the shared state. 
Ideally, if the triggered features are not added by the adversary, the attack remains benign in most cases, and the victim will be unaware that their private location has been manipulated by the attacker. 
\subsection{Evaluation}
\label{sec:cloudanchor-eval}

Next, we evaluate the attacker's success rate in both the remote write, remote read, and triggered remote write attacks in different environments, including investigating the impact of clutter, lighting, and indoor vs. outdoor environments.

\subsubsection{Remote Read Evaluation}
\label{sec:remote-read-eval}

\begin{table}[ht]
\centering
\small
\begin{tabular}{|C{4cm}|C{1.5cm}C{1.5cm}|}
\hline
\multirow{2}{*}{\textbf{Environment}} & \multicolumn{2}{c|}{\textbf{Attack success rate}}      \\ \cline{2-3} 
                                       & \multicolumn{1}{c||}{Static scene} & Add clutter \\ \hline \hline
Office desk                            & \multicolumn{1}{c||}{13/16}          & 10/16      \\ \hline
Bedroom desk                           & \multicolumn{1}{c||}{12/16}          & 7/16      \\ \hline
Bedroom bed                            & \multicolumn{1}{c||}{14/16}          & 7/16      \\ \hline
Outdoor garden                            & \multicolumn{1}{c||}{5/16}          & 2/16      \\ \hline
Outdoor BBQ                              & \multicolumn{1}{c||}{16/16}          & 15/16      \\ \hline
Outdoor pool                             & \multicolumn{1}{c||}{16/16}          & 15/16      \\ \hline
\end{tabular}
    \caption{Success rates of remote read attacks in \textit{Static scene} and \textit{Add clutter} conditions. Attacks succeed often and perform better in a \textit{Static scene} compared to \textit{Add clutter}.}
    \label{ttable3}
\end{table}

\paragraph{Experiment setup.} 
We execute the remote read attack in six different environments, as shown in Table~\ref{ttable3}.
These environments include a range of backgrounds, including an office, personal home, and pool, with about half being indoor environments and the other half outdoor (see Fig.~\ref{fig:arcore_environments} in the Appendix).
All of the experiments were been done with a Samsung Galaxy S20 Android phone, and an Apple MacBook Pro served as the monitor to display the environment images.
The success rate was used as the evaluation metric. 
It was defined as the number of trials in which we were able to successfully read the written hologram remotely. 
\revision{We also evaluated a benign baseline where the user reads the hologram from the real environment.}

Furthermore, we evaluated the success rate under two conditions: \textit{Static scene} and \textit{Add clutter}.
The motivation for studying these two conditions is to simulate the case where the attacker does not have perfect information about the victim's environment or the environment has changed in the interim.
 In the \textit{Static scene} condition, the victim's true environment closely resembles the attacker's image of the environment.  The \textit{Add clutter} condition involves environments that have new objects or alterations in the attacker's image compared to the victim's original environment. 
It is a more challenging condition because there are additional features during the attacker's read process which were not been present during the victim's write, \ie the read key may not exactly match the write key, so the attacker's read may fail. 
 The results of these evaluations provide insights into the effectiveness of the remote read attack under different conditions.

\begin{figure}
  \begin{center}
    \includegraphics[width=0.40\textwidth]{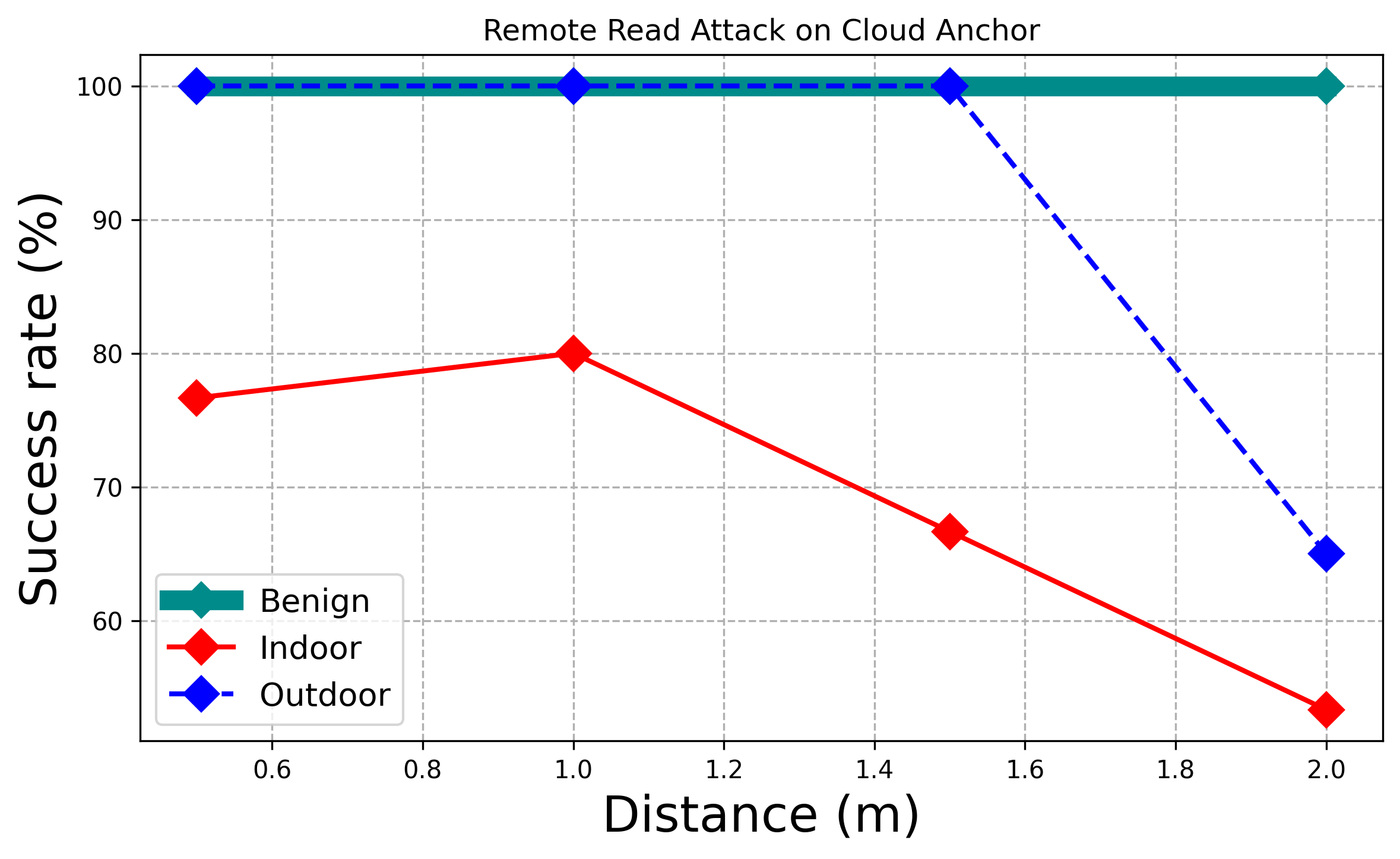}
  \end{center}
  \vspace{-0.2in}
  \caption{\revision{Remote read attacks in Scenario A.}}
  \label{fig:RR_distance_result}
\end{figure}

\paragraph{Results.}
As Table~\ref{ttable3} shows, the success rate of the attack is generally good, with the attack succeeding about half the time, on average, across all of the environments we experimented in.
This makes sense because, according to our observations, the critical phase of shared state communications is usually the writing process. The better the quality of the key uploaded by the victim during her write request, the easier it is for subsequent users (including the attacker) to read successfully. 
In other words, because the writing was performed in the real environment (not from a photograph) by the victim, there are many 3D features extracted from the victim's camera images and inserted into the map in the shared state.
This creates a larger attack surface because there are many possible matching keys (\eg different angles of the scene) that an attacker could use to successfully launch a remote read.

\revision{We conducted additional experiments at different distances between the AR device's camera and the attacker's image of the environment.
The results  in Fig.~\ref{fig:RR_distance_result} show that the attack success decreases as the attacker moves further away and the image becomes smaller in the camera's field of view.
On the other hand, if the distance is less than 0.5 meters, it becomes challenging for the camera to focus on the visual features of the image, and the attack tends to be less successful.
The benign baseline can read the hologram the vast majority of the time, while the attacker is only moderately less successful. 
}

\subsubsection{Remote Write Evaluation}
\label{sec:cloudanchor-write-eval}

\paragraph{Experiment setup.} 
The setup is similar to the remote read attack (\S \ref{sec:remote-read-eval}). 
A minor difference is that the \textit{Add clutter} condition refers to environments where there are additional objects or changes in the victim's real environment (during the read) compared to the attacker's image (used to do the poisoned write).
We also informally experimented with different environment lighting, conducting experiments in both brightly lit and dimmer versions of the same environment (\eg by turning on/off a lamp or daytime/sunset).

\begin{table}[t]
\small
\begin{tabular}{|C{4cm}|C{1.5cm}C{1.5cm}|}
\hline
\multirow{2}{*}{\textbf{Environment}} & \multicolumn{2}{c|}{\textbf{Attack success rate}}      \\ \cline{2-3} 
                                       & \multicolumn{1}{c||}{Static scene} & Add clutter \\ \hline \hline
Office desk                            & \multicolumn{1}{c||}{8/16}          & 7/16      \\ \hline
Bedroom desk                           & \multicolumn{1}{c||}{6/16}          & 4/16      \\ \hline
Bedroom bed                            & \multicolumn{1}{c||}{10/16}          & 8/16      \\ \hline
Outdoor garden                             & \multicolumn{1}{c||}{1/16}          & 0/16      \\ \hline
Outdoor BBQ                              & \multicolumn{1}{c||}{16/16}          & 15/16      \\ \hline
Outdoor pool                             & \multicolumn{1}{c||}{15/16}          & 14/16      \\ \hline
\end{tabular}
    \caption{Success rates of remote write attacks in \textit{Static scene} and \textit{Add clutter} conditions. The overall success rate of remote write attacks is slightly lower than that of remote read attacks. The success rates decrease in the \textit{Add clutter} condition.       
    }
    \label{tab:remote-write-clutter}
\end{table}

\paragraph{Results.}



Table~\ref{tab:remote-write-clutter} shows the success rates of the remote write attack in different environments. As can be seen, the attack reaches a high degree of success in both indoor and outdoor environments.
The success rate is generally lower than the remote read attack because, as discussed earlier in Section~\ref{sec:cloudanchor-method}, a write request generally requires more camera images in its key, and for an attacker to capture these multiple camera images from different angles of a single photograph is challenging.
The ``Outdoor garden'' scene has a particularly low success rate. This can be attributed to the limited number of planes present in the scene compared to other environments, 
as the API typically relies on an adequate number of planes to create a map in the shared state and enable writing. However, it is important to emphasize that the low success in this environment affects both the remote write attack and a legitimate write process equally. 
\revision{We also evaluated the impact of distance on the attack, and the results follow those of the indoor scenario (presented in Fig.~\ref{fig:RW_distance_result} in Appendix~\ref{app:cloudanchor}).}

Our attack demonstrates strong robustness against environmental factors like lighting changes, clutter, and \revision{distance}, which barely affect its success rate. Based on our experiments, we have observed that when the actual environment that the victim is in is significantly darker than what is shown in the attacker's image, such as during nighttime or when the lights are almost turned off, the success rate of the attack degrades by approximately 15-25\%.
The robustness of our attack, which doesn't require precise knowledge of the victim's environment, makes it particularly dangerous. Notably, clutter impacts the success rate of remote read attacks more than remote write attacks. This occurs because remote reads face two layers of noise (photographic noise and clutter) simultaneously, reducing success rates. However, for remote write attacks, the layers of noise are separated (the photograph adds noise during the attacker's write, and the added clutter adds noise during the victim's read), and thus the impact on the success rate is less.
 


\subsubsection{Triggered Remote Write Evaluation}
\label{sec:trigger_remote_read_eval}
\paragraph{Experiment setup.} The setup is similar to the remote write experiment in the previous attack (Section~\ref{sec:cloudanchor-write-eval}), except that the attacker adds additional trigger features during remote write as depicted in Fig.~\ref{fig:trigger-attack}. 
For our experiments specifically, we have used a simple piece of paper with some marks on it and a spinner on the paper placed near the image on the monitor.
During the attacker's remote write, we do our best to move the attacker's camera to capture features both from the image on the monitor and the additional trigger features. 
In addition to having the victim read from the same environment as the attacker's write, we also examined the false positive rate in two cases: (case 1A) whether the victim can view the hologram in a different environment containing the trigger and (case 1B), whether the victim can view the hologram in the correct environment without the trigger present.
Ideally, the false positive rate should be low in both cases.

\begin{table}[t]
\small
\begin{tabular}{|C{4cm}|C{1.5cm}C{1.5cm}|}
\hline
\multirow{2}{*}{\textbf{Environment}} & \multicolumn{2}{c|}{\textbf{Attack success rate}}      \\ \cline{2-3} 
                                       & \multicolumn{1}{c||}{Static scene} & Add clutter \\ \hline \hline
Office desk                            & \multicolumn{1}{c||}{15/16}          & 15/16      \\ \hline
Bedroom desk                           & \multicolumn{1}{c||}{13/16}          & 12/16      \\ \hline
Bedroom bed                            & \multicolumn{1}{c||}{15/16}          & 13/16      \\ \hline
Outdoor garden                              & \multicolumn{1}{c||}{3/16}          & 1/16      \\ \hline
Outdoor BBQ                              & \multicolumn{1}{c||}{16/16}          & 16/16      \\ \hline
Outdoor pool                             & \multicolumn{1}{c||}{16/16}          & 16/16      \\ \hline
\end{tabular}
    \caption{Success rates of triggered remote write attacks in \textit{Static scene} and \textit{Add clutter} conditions \textbf{with triggered features}. The success rates are nearly identical in the \textit{Static scene} and \textit{Add clutter} conditions.}
    
    
    \label{ttable2}
\end{table}

\paragraph{Results.} Table~\ref{ttable2} shows the results derived from the experiments. As the results suggest, there is a large boost in the success rate compared to the vanilla remote read attack results in Table~\ref{tab:remote-write-clutter}.
We examined two critical aspects of the triggered remote write attack: the false positive rate in cases 1A and 1B.
Fortunately, in case 1A, false positives never happen in our experiments; we believe this is probably because the trigger features that we used are very simple and constitute a relatively small fraction of features from the entire environment.
In other words, they act as auxiliary features and are not sufficient alone 
for the victim to use them as a key to read the hologram.
In case 1B, the victim can sometimes (around 50\% of the time) still read the hologram even if the trigger used by the attacker during the remote write is absent from the scene. 
This aligns with our hypothesis that the trigger features serve as auxiliary features in the scene. 
These results are comparable to the those achieved by Ji et al.~\cite{poltergeist} without adversarial patch triggers, in a non-AR domain. However, we find that if an adversary adds the triggered features to the victim's environment, the victim will read the attacker's hologram from the shared state with a much higher success rate of over 90\%, so the triggers are effective.

\section{Scenario B: Global, Curated Shared State}
\label{sec:geospatial}

Built on Google's extensive database of public street images, the Geospatial API~\cite{Geospatial} allows users to attach AR holograms to any location within Google Street View, creating an AR experience on a global scale.
This is an example of a global, curated, shared state.
In this section, we demonstrate a practical attack in which the attacker can remotely read to steal a private hologram written by the victim.
For example, in a city-wide scavenger hunt, an attacker could cheat to collect the virtual treasure simply by trying images of the possible treasure locations.
\revision{The attacker benefits by eliminating the physical labor needed to visit all the locations, gaining a competitive advantage over other users.}
The attack technique is similar to those on the local, curated shared state discussed in Section~\ref{sec:cloudanchor}, but the main difference is the addition of GPS as a key (along with camera images), which requires changes to the attack methodology. Also, the Geospatial API, being built on Google Street View, limits the read and write of holograms to outdoor environments. However, we have discovered that by manipulating GPS, camera, and IMU readings, we are able to deploy remote read attacks indoors as well.
\revision{All experiments were performed on local applications and devices only accessible by us, without malicious writes to the shared state, so they did not harm external users.}
\subsection{Methodology}


The Geospatial API gives users the capability to place holograms in their physical surroundings by leveraging spatial data obtained from Google's Visual Positioning System (VPS)~\cite{GoogleVPS}, based on Street View images. Using computer vision algorithms on the camera images, the API facilitates the accurate determination of the device's location and orientation to locate and display the correct holograms, surpassing the localization capabilities of GPS alone. However, this technology also introduces potential security vulnerabilities that can be exploited by malicious actors.

\begin{figure}
  \begin{center}
    \includegraphics[width=0.42\textwidth]{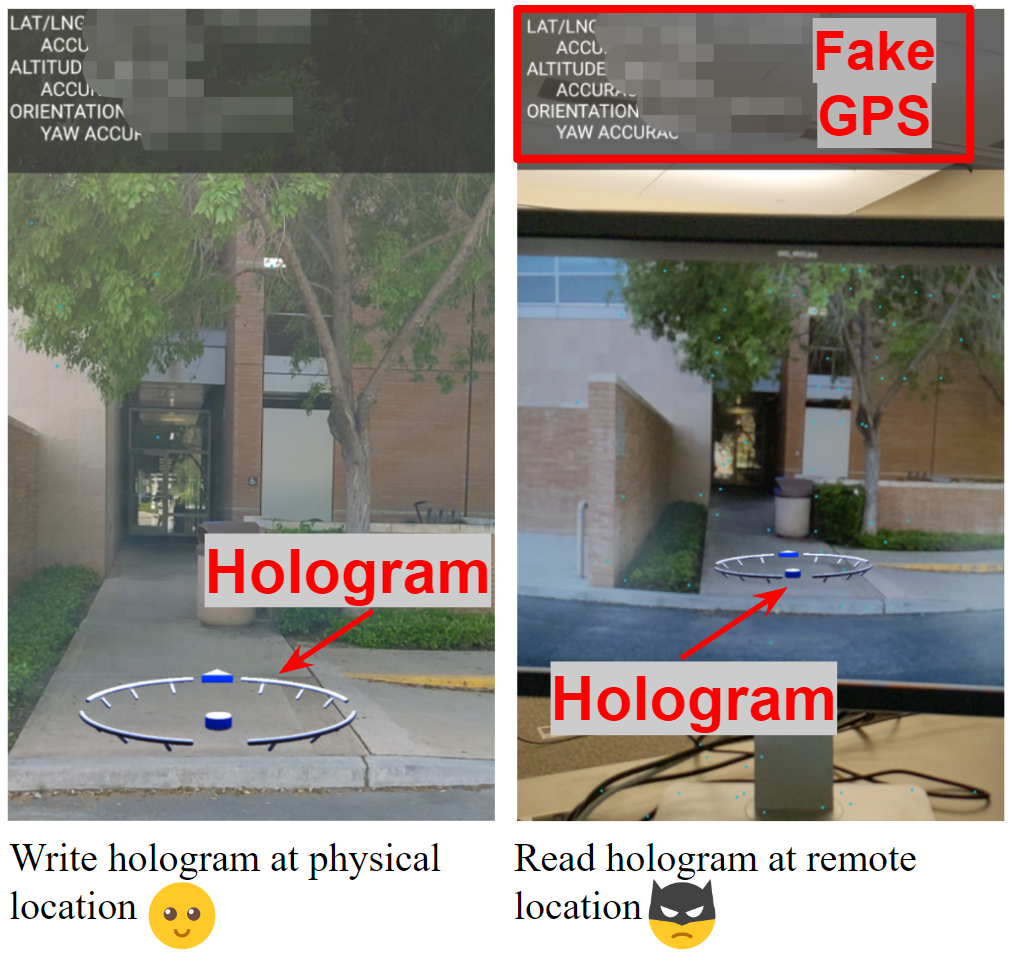}
  \end{center}
  \vspace{-0.2in}
  \caption{Remote read attack in Scenario B. \emph{Left:} A legitimate user places a hologram in front of a building. \emph{Right:} The attacker can view the hologram on an image of the building. 
  }
  \label{fig:geospatial-remote-read}
\end{figure}

\paragraph{Remote read attack.} By employing GPS spoofing applications, an attacker can remotely read holograms by altering the GPS location of her device. 
Along with utilizing a GPS emulator, the attacker points her device's camera toward printed photographs or virtual images displayed on a monitor in order to generate a poisoned read request to the shared state and view the hologram at the target location.
\revision{We assume the attacker has sourced these} photographs/images from public online platforms, such as Google Street View, 
Mapillary~\cite{neuhold2017mapillary}, or even real estate websites.
Fig.~\ref{fig:geospatial-remote-read} demonstrates the process, illustrating how the attacker successfully manipulates the device's GPS location using a GPS emulator \revision{combined with image spoofing} to achieve the remote reading of holograms onto her AR display. 

\subsection{Evaluation}
\label{sec:geo-evaluation}

\paragraph{Experiment setup.}
To begin with, we place 23 holograms at various campus locations 
using the Geospatial API. 
We selected these locations to encompass a range of environmental differences and varying light conditions, \revision{shown in Fig.~\ref{fig:geospatialexample} in the Appendix}. Subsequently, we capture photographs of the areas where the holograms were placed. 
We employ a GPS emulator application~\cite{GPSEmulator} to generate fake GPS locations on the Android phones utilized for testing. By manipulating the GPS coordinates and displaying an image of the target location,  we aim to deceive the shared state into returning the associated holograms at those locations.


 We conducted the remote read attack with the attacker's device located from [0.25, 0.5, 0.75, 1, 1.5, 2] meters away from the monitor.
To assess the effectiveness of these attacks, we utilize the attack success rate as the primary metric. We define a successful attack when each read operation can succeed in less than three trials. Our testing involved two Android phones, namely the Samsung Galaxy S8 and the Samsung Galaxy S21. The former was used by the victim to place the holograms, while the latter was used by the attacker 
to capture the images (size 3024 x 4032 pixels) and conduct the attacks. 

\begin{figure}
  \begin{center}
    \includegraphics[width=0.40\textwidth]{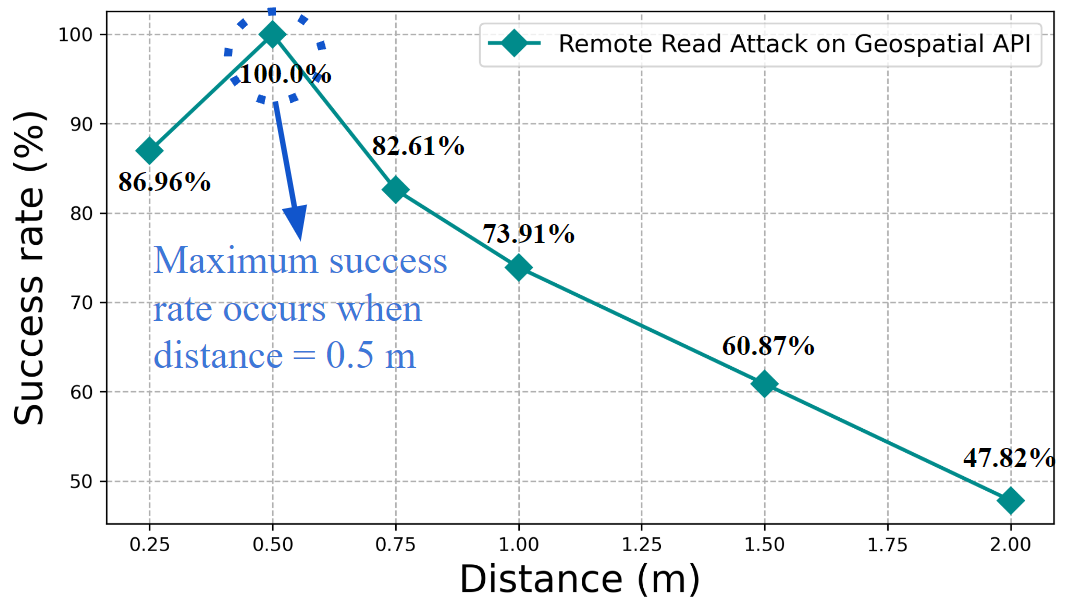}
  \end{center}
  \vspace{-0.2in}
  \caption{Remote read attacks in scenario B. 
  }
  \label{fig:distance-result}
\end{figure}

\paragraph{Results.} Fig.~\ref{fig:distance-result} shows our findings in terms of the attack success rate as a function of the attacker's distance.
When the distance between the attacker's device and the monitor is too close, such as at 0.25 meters, the camera on the device may struggle to focus properly. This can result in blurred images, making conducting successful remote read attacks challenging. Notably, we achieved a 100\% success rate for remote read attacks conducted at a distance of 0.5 meters. This distance proves to be optimal for the camera on the device to focus properly, resulting in clear and discernible images. However, as the distance between the attacker's device and the monitor increases, the success rate of the remote read attacks declines significantly. We speculate that several factors may influence this decline in success rate. Firstly, as the distance increases, the images displayed on the monitor become smaller, making the attacker's read key significantly different from the victim's initial write key.
Similarly, when the device is positioned at a greater distance from the monitor, there is an increased likelihood of capturing unrelated objects in the field of view. This can significantly impact the success rate of the attack.
\revision{In contrast, the benign baseline, without any attack in the real environment, succeeds 100\% of the time as expected, but our attack still succeeds most of the time.}


\section{Scenario C: Global, Crowd-Sourced Shared State}
\label{sec:mapillary}

\label{sec:GlobalCrowd}
    While the imagery needed for global AR exists in Geospatial Anchor, as discussed in the previous section, such services' shared states are curated, meaning that only trusted individuals (paid contractors) are able to gather and upload data to write to the map. However, more recent services like Mapillary~\cite{Mapillary} allow users to both read from the map and write new data to expand and update it.
    Mapillary is a non-curated service, which means that all users have the ability to read and write in the shared state. This includes the raw map data as well as holograms that are virtual representations of real objects (\eg traffic signs, fire hydrants, and light poles).

    Non-curated applications that rely on GPS and camera images as keys, such as Mapillary, introduce new attack vectors, as attackers with minimal permissions gain more capabilities.
     \revision{Although allowing users to read or write to the shared state is the desired behavior of a crowd-sourced platform, carefully poisoning the read or write updates can cause adverse downstream effects that are unique to AR.}
    Towards this, in this section, we investigate attacks on global, crowd-sourced AR shared states, using Mapillary as an example.
    We investigate two types of attacks: a poisoned write to the map in the shared state (Section~\ref{sec:mapillaryswap}), and a poisoned write that creates false holograms in the shared state (Section~\ref{sec:mapillaryfake}).
    \revision{For the former, we will demonstrate the construction site example from \S\ref{sec:intro}, where the attacker causes a victim worker to see the wrong construction signs and dig in unsafe real-world areas.}

    All experiments conducted in this section were carried out with permission from Mapillary. The experiments were conducted within a designated geo-fenced area, which was specifically created for the purpose of these experiments. 
    \revision{This, in addition to Mapillary-provided test accounts, ensured that data uploaded for the experiments only appeared on the test accounts and devices, and had no impact on external users.
    }

\subsection{Poisoned Write to the Shared State's Map}
\label{sec:mapillaryswap}


\subsubsection{Methodology}

The high-level idea is for the attacker to poison the GPS part of the key associated with the write request while keeping the hologram part of the key legitimate.
Specifically, the attacker obtains image sequence A and image sequence B from two locations, A and B.
Normally, these locations should be associated with holograms A and B, respectively.
It swaps their GPS coordinates and makes two write requests: one with \{key=image sequence A + GPS B, value=hologram B\}, and another with \{key=image sequence B + GPS A, value=hologram A\}.
Thus the hologram at location B becomes associated with image sequence A, and vice versa.
Because of this, a victim who later reads with \{key=image sequence A\} will receive a response from the shared state with \{value=hologram B\}, and view the wrong hologram at location A.

Note that while the mechanics of this attack are similar to the global, non-curated attack (Section~\ref{sec:geospatial}) in that the GPS part of the key is modified, here we focus on write attacks rather than read attacks, which means that we need to carefully craft the spoofed GPS (by swapping) during the write request, in order to cause adverse effects to downstream victims who read the poisoned shared state.
Next, we describe the detailed attack mechanics in terms of writes and reads.


\begin{figure}
  \begin{center}
    \includegraphics[width=0.40\textwidth]{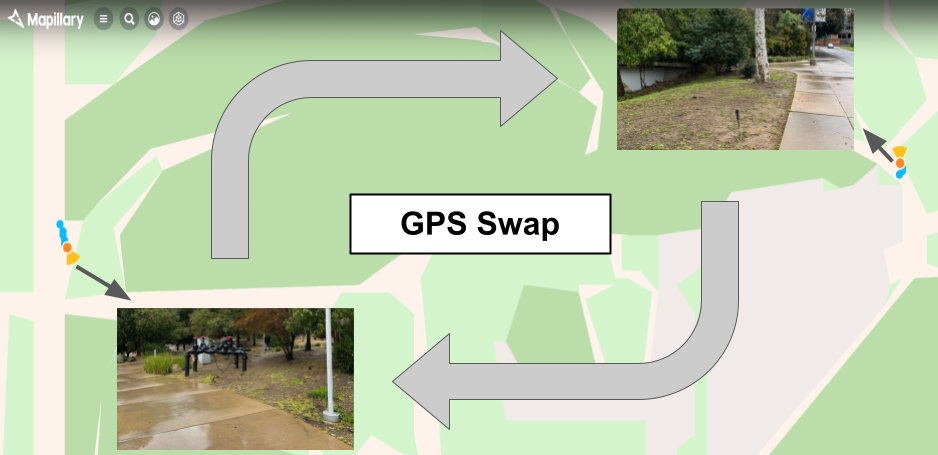}
  \end{center}
   \vspace{-0.2in}
  \caption{Two image sequences with their GPS coordinates swapped are shown in a screenshot of the Mapillary shared state map.}
  \label{fig:GPSswapExample}
\end{figure}

\paragraph{Attacker's write mechanics.}
To upload images to the shared state, a sequence of images, each with associated metadata (latitude, longitude, altitude, and time), is needed. The sequence can range from a minimum of three images to several hundred.
All of the image metadata is in Exchangeable Image File (EXIF) format, whose data is
freely manipulable using scripts.
An attacker can modify the metadata so that the image seems to have been captured at any arbitrary location and time (within reason; for example, timestamps from the future will be rejected by the shared state).
An illustration of image sequences with modified metadata being successfully ingested by the shared state is shown in Fig.~\ref{fig:GPSswapExample}.
In this particular example, two sequences of five images each, captured using an iPhone 12, were uploaded with swapped latitude, longitude, altitude, and time.
The images were successfully uploaded using Mapillary's desktop uploader utility~\cite{mapillary-uploader}.
Mapillary allows these sequences to be uploaded, processed, and displayed at the swapped locations for viewing by other users.



\begin{figure}
    \centering
    \includegraphics[width=0.42\textwidth]{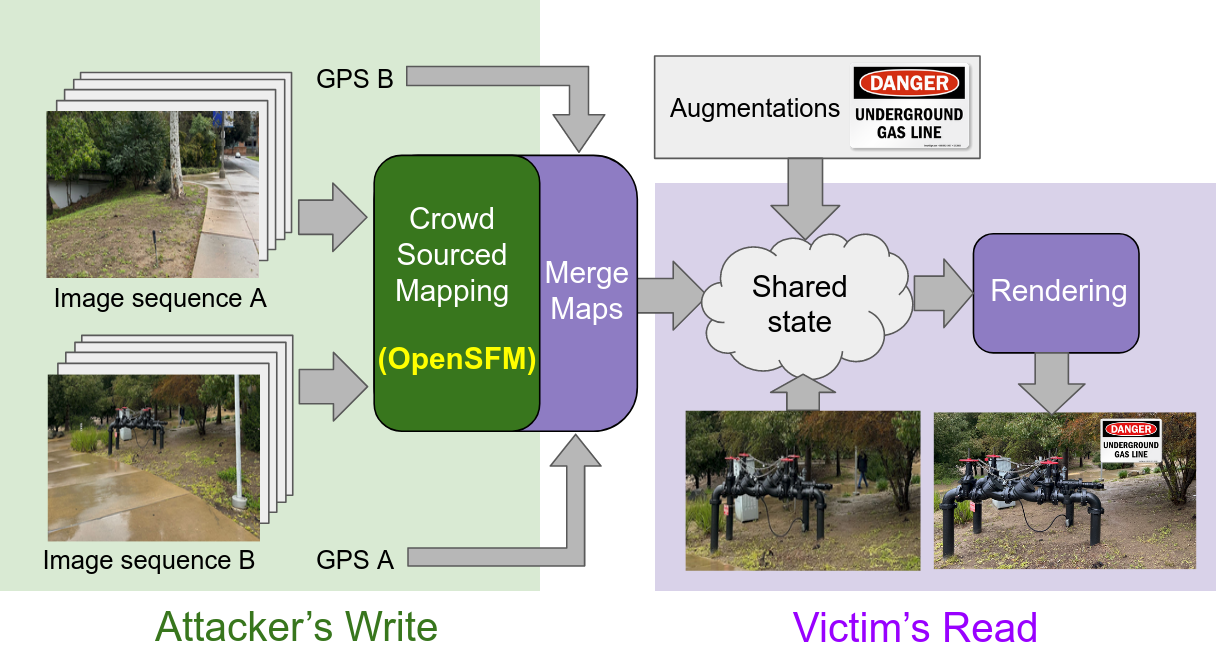}
 \vspace{-0.1in}
    \caption{Details of poisoned write to the shared state's map mechanics. We used Mapillary's open-source OpenSFM~\cite{OpenSFM} library to demonstrate the attack with swapped GPS keys. 
    }
    \label{fig:ararchitecture}
\end{figure}

\paragraph{Victim's read mechanics.}
\revision{After the attacker's poisoned write on Mapillary, we next turn our attention to visualizing the results on the victim's side.}
One challenge that we faced is that Mapillary is a closed source and does not currently have a public AR interface to experiment with, which is needed in our study to read the shared state and demonstrate the impact on AR victims.
\revision{To overcome this, we utilize Mapillary's underlying computer vision library, OpenSFM~\cite{OpenSFM},
    plus our own additional Python scripts to construct a simple AR viewer that replicates, to the best of our ability, how a legitimate user's read request would be visualized.}
    Fig.~\ref{fig:ararchitecture} shows the data flow of the AR application from an image sequence to the hologram.
    First, the initial maps in the shared state are generated using OpenSFM~\cite{OpenSFM} from the data in the attacker's write request.
    Next, the spoofed GPS from the write request is used to facilitate the processing of the maps that are then stored in the shared state alongside the augmentations.
    Finally, a victim captures new images and uploads them to the shared state in a read request, which is processed by the OpenSFM to return nearby holograms for rendering on the victim's display.
    This step is similar to how a commercial AR service would handle read requests~\cite{schmalstieg2016augmented,MultiConsistent}.
    We do not use GPS as a key during the read because OpenSFM does not support it, and frameworks that support image and GPS keys are research prototypes only~\cite{boche2022visual}.

\subsubsection{Evaluation}


We repeated the write attack a total of 8 times on Mapillary's shared state for 15 total image sequences with false GPS data (one sequence was a duplicate, not a swap). 
These swaps occurred using imagery captured outdoors within 1 km$^2$ of the geo-fenced area.
The images were taken at different times of day, ranging from early morning to early evening, facing different directions, and at different locations (\eg streets, and grass fields without roads).
\revision{These images were captured on iPhone 12 at 1080p and uploaded in PNG format.}
We verified through the Mapillary web interface that all the attacker's write requests with spoofed GPS data were successfully ingested by the Mapillary pipeline, uploading, processing, and displaying the spoofed imagery.
This shows that the fundamental write attack mechanic works.
The main reason this works is that while Mapillary does check for basic undesirable content, it does not check whether crowd-sourced images indeed correspond to the claimed GPS locations. 

\begin{figure}[t]
    \centering
    \resizebox{0.46\textwidth}{!}{%
    \begin{tabular}{ccc}
     & Location A & Location B \\
    \begin{turn}{45}No Attack \end{turn} &
    \includegraphics[width=0.17\textwidth]{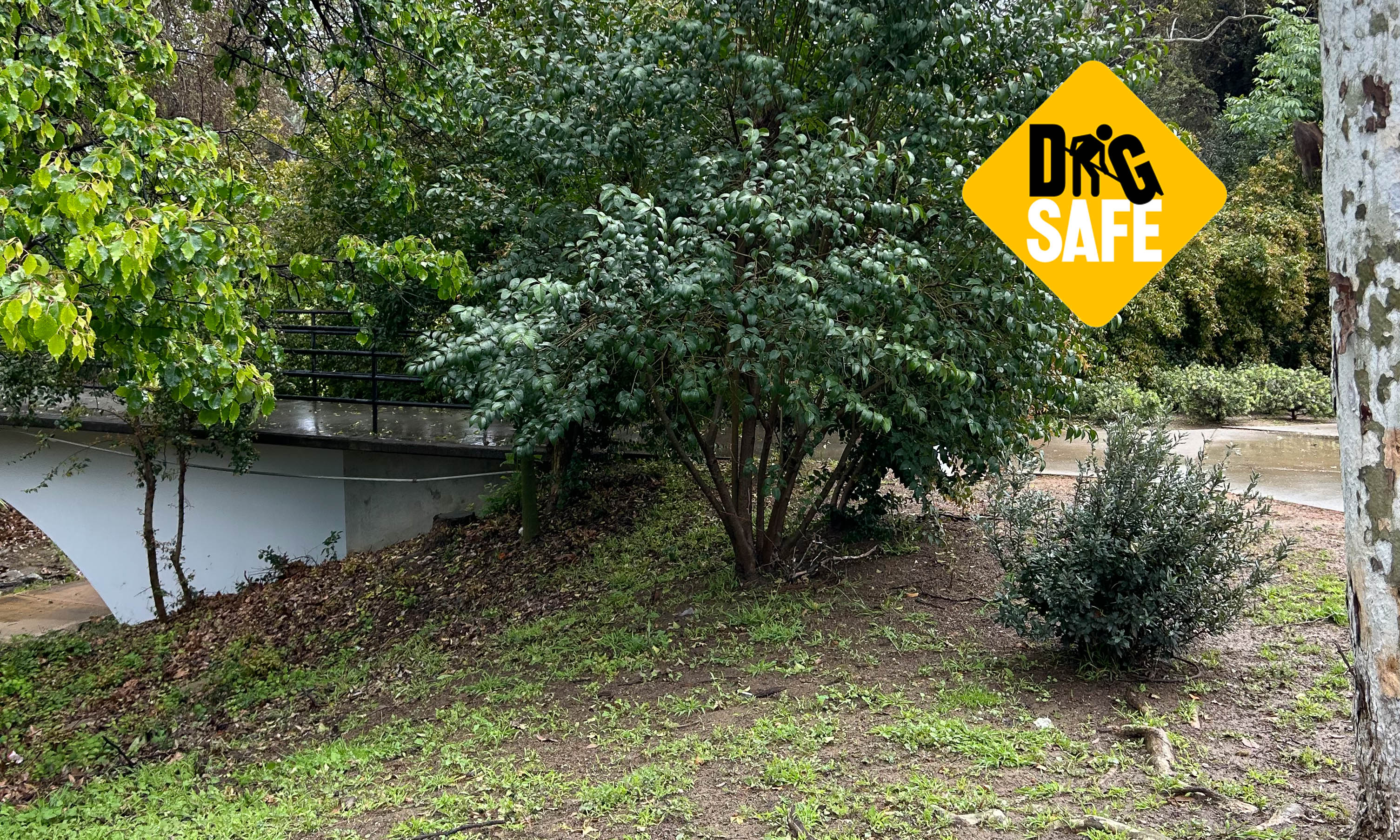} &
    \includegraphics[width=0.17\textwidth]{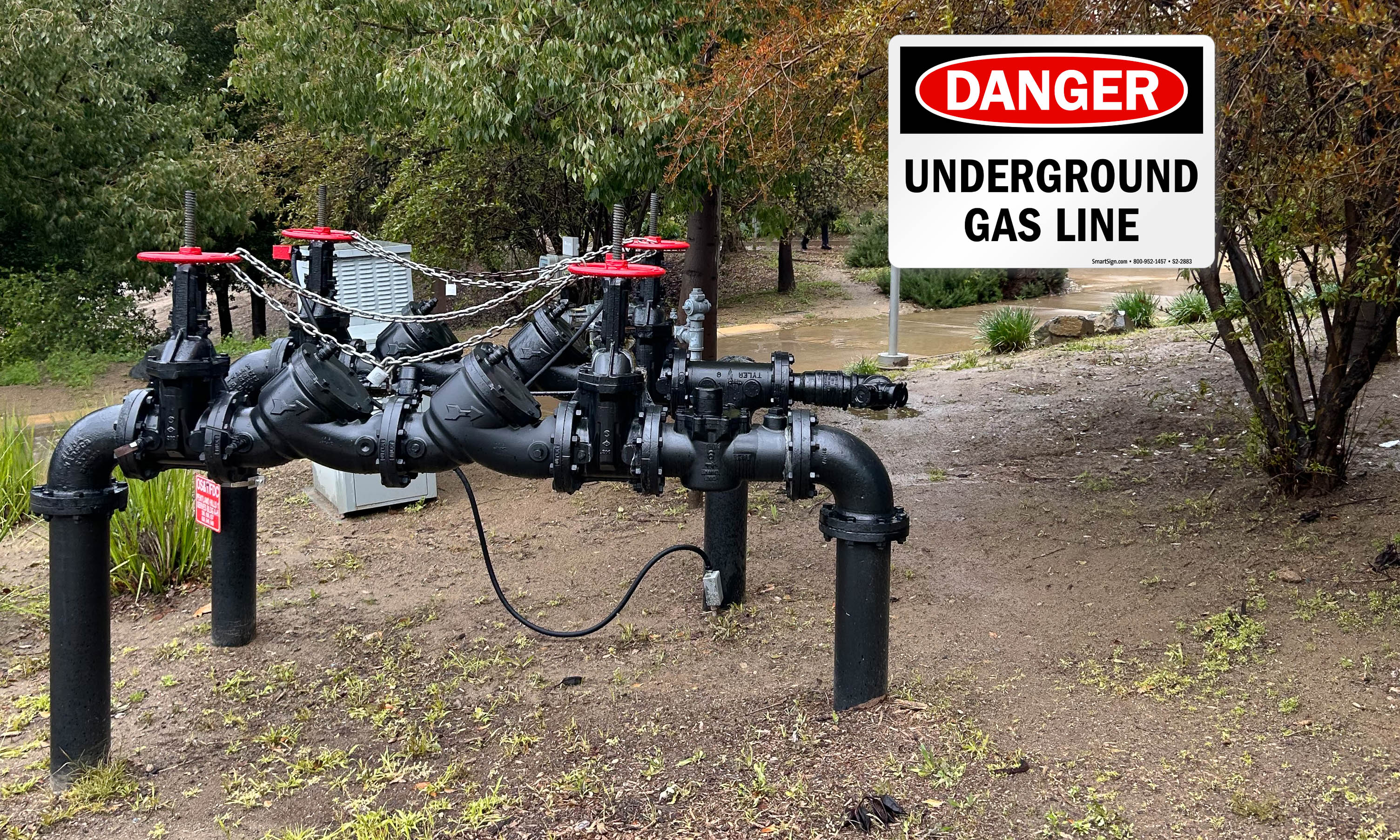} \\
    \begin{turn}{45}With Attack \end{turn} &
    \includegraphics[width=0.17\textwidth]{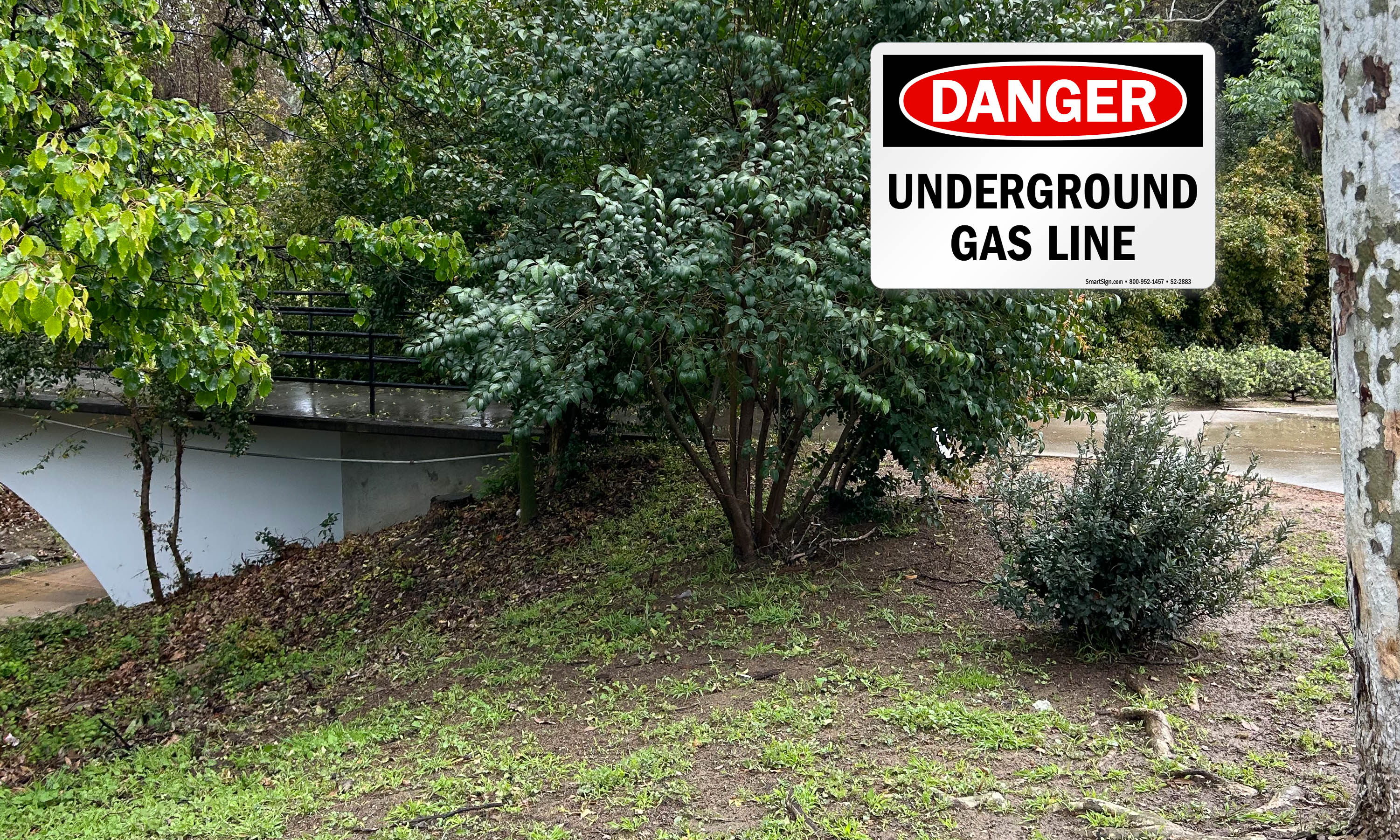} &
    \includegraphics[width=0.17\textwidth]{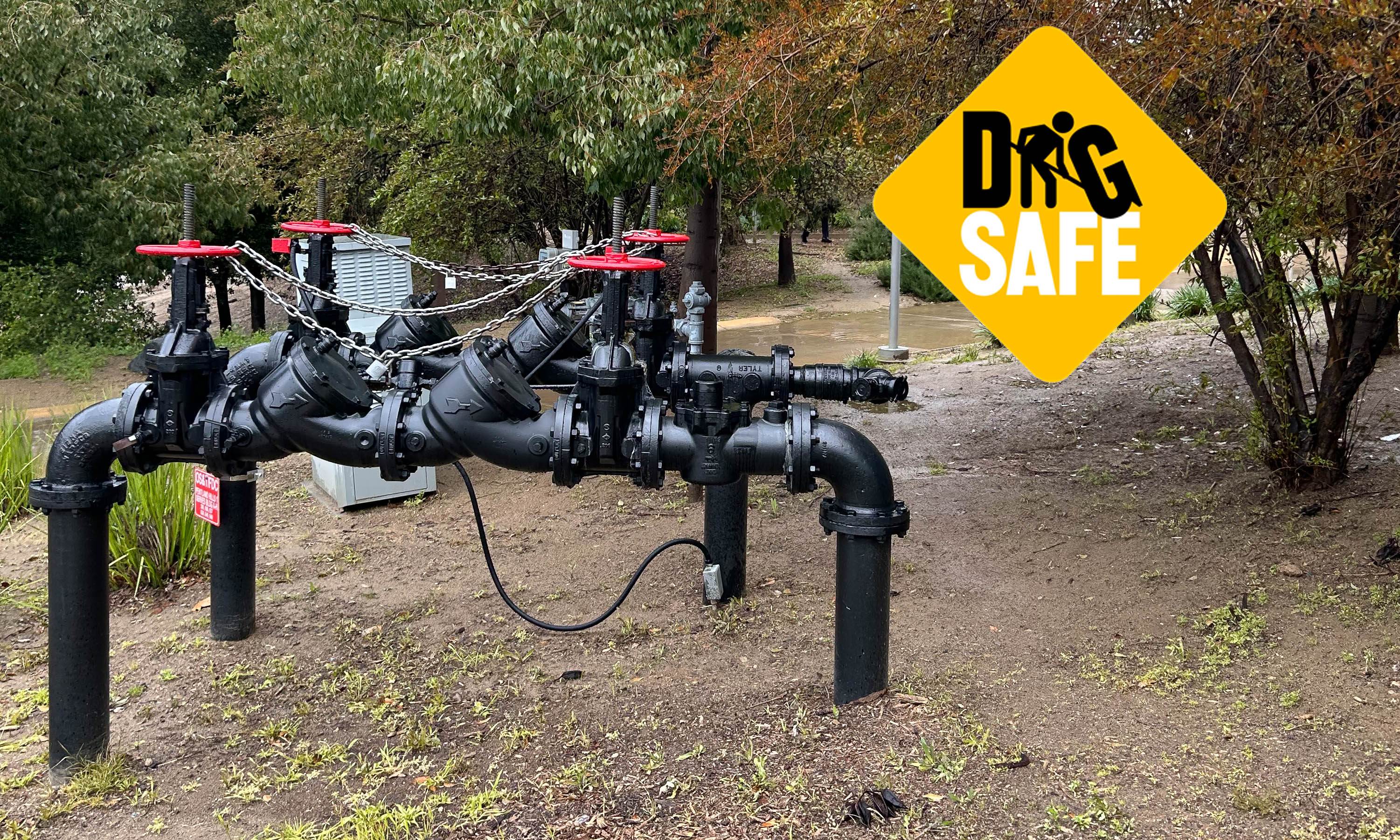} \\
    
    
    \end{tabular}%
    }
    \caption{Effect of poisoned write to the shared state's map. Holograms are read at the wrong locations; in this example, a ``safe to dig'' sign is placed next to an underground pipe.}
    \label{fig: ARSwap}
\end{figure}

With the write attack mechanics validated on Mapillary, we next sought to show the impact on a victim AR user. 
To showcase this, we had the attacker write two image sequences (a grass scene and a pipe scene), containing five images each, to the shared state.
We reserved one additional image per sequence for use by the victim.
The holograms (a ``dig safe'' sign and a ``danger: underground gas line'' sign) included in the write request were associated with locations 5 meters in front of the first image in each sequence.
After running through the pipeline in Fig.~\ref{fig:ararchitecture}, the final display to the victim is shown in Fig.~\ref{fig: ARSwap}.
The first row shows the AR display seen by a victim without our attack.
The ``dig safe'' hologram is displayed in the grass field, and the ``danger'' hologram is displayed near the pipes, as intended.
The bottom row shows that with our attack, the wrong hologram (``dig safe'') is shown near an underground gas line, leading to serious safety issues.

\subsection{Poisoned Write of Shared Holograms} 
\label{sec:mapillaryfake}

In this subsection, we discuss another vulnerability through modifications to the image sequence part of the key.
Some AR shared states (\eg Mapillary) perform object detection on the images uploaded by users to their service~\cite{MapObj}. When an image sequence is uploaded, these detected objects are added to the shared state map at the positions they were detected.
This presents attackers with the opportunity to tamper with the images and introduce fake holograms into the shared state.
For example, the attacker could create a fake stop sign hologram overlaid onto an otherwise empty street, causing an AR navigation app to provide wrong directions to the user.

\paragraph{Methodology.}
We used Photoshop to edit a sequence of images to add a stop sign and write them to the shared state.
For the attack to be successful, the stop sign's size had to be proportional to the user's distance from it, and the octagonal shape preserved using transparency layers. 
Mapillary required photo-realistic stop signs in order for the fakes to be successfully ingested and recognized.
The fake object also had to be present in at least 3 images in order for Mapillary to place it accurately, which requires multiple photographs of the stop sign with appropriate scaling.
Without these changes, the Mapillary pipeline rejected the write request.

\begin{figure}[t]
    \centering
     \begin{subfigure}[b]{0.23\textwidth}
      \includegraphics[width=\textwidth]{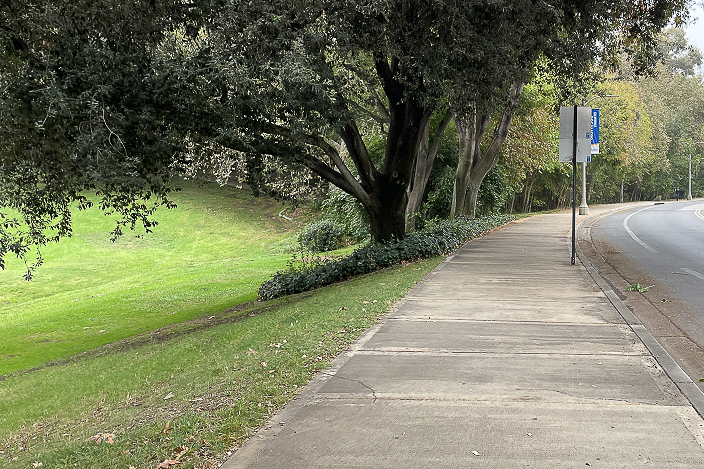}
      \caption{Real-world ground truth.}
      \label{fig:stopFake-sub1}
    \end{subfigure} 
    ~
     \begin{subfigure}[b]{0.23\textwidth}
      \includegraphics[width=\textwidth]{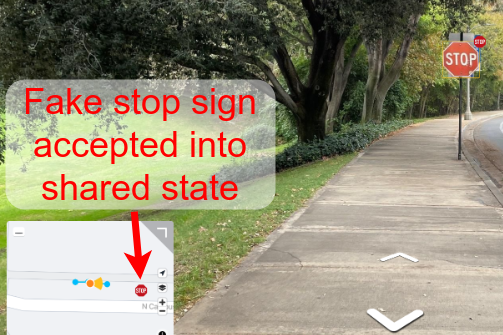}
      \subcaption{Tampered image.  }
      \label{fig:stopFake-sub2}
    \end{subfigure}
    \caption{Poisoned write to the shared state's holograms. A fake stop sign has been inserted into a sequence of images in order to fool the shared state's object detector, resulting in a fake stop sign hologram being added to the shared state. }
    \label{fig:StopFake}
\end{figure}

\paragraph{Evaluation.}
Fig.~\ref{fig:StopFake} shows an example of a successful attack.
Fig.~\ref{fig:stopFake-sub1} shows the real-world ground truth.
Fig.~\ref{fig:stopFake-sub2} shows the tampered image, with a photograph of a stop sign taken from public sources cropped and overlaid on top of it.
The small subfigure in the bottom left of Fig.~\ref{fig:stopFake-sub2} shows a screenshot of Mapillary's web interface where the stop sign hologram is accepted into Mapillary's shared state. 
We expect that attackers could also write other false holograms into the shared state; any of Mapillary's pre-defined object detection classes (\eg traffic signs, lamp posts) could work.


\section{Shared State Attack Mitigations} 
\label{sec:mitigations}

The fundamental question at issue for these attacks is how to accurately establish the true location of an AR device.
All of these attacks involve deceiving the shared state about the attacker's location to read or write data maliciously.
We discuss multiple potential mitigation strategies related to this \revision{and our defense prototype utilizing additional sensor modalities.}

\begin{figure}[t]
    \centering
    \begin{subfigure}[b]{0.230\textwidth}
    \includegraphics[width=\textwidth]{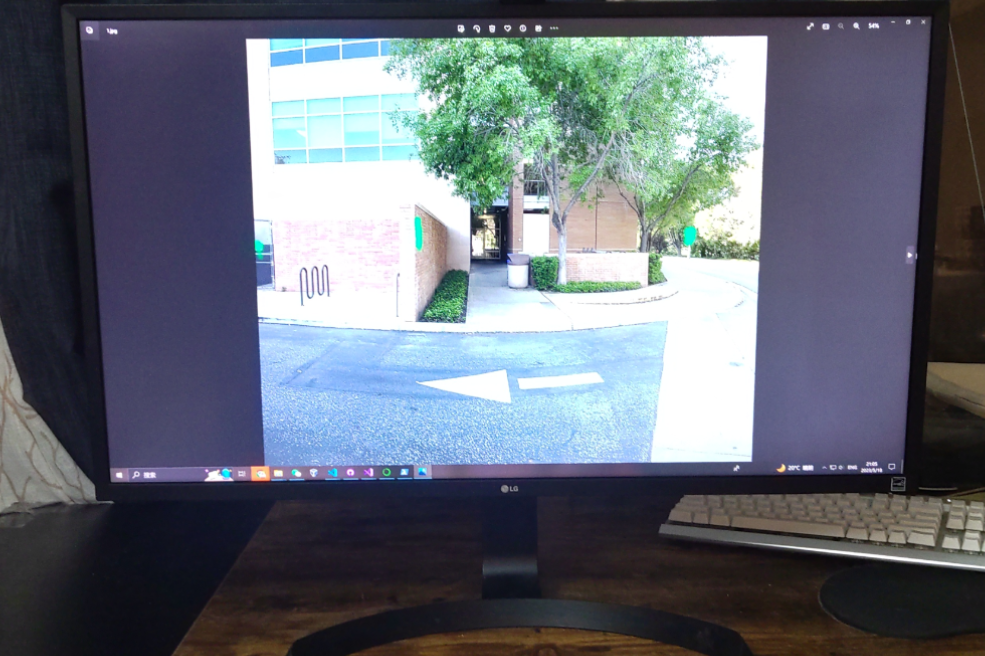}
    \caption{Hololens 2 RGB camera.}
    \label{fig:hololens_defense-sub1}
    \end{subfigure}
    ~
    \begin{subfigure}[b]{0.230\textwidth}
    \includegraphics[width=\textwidth]{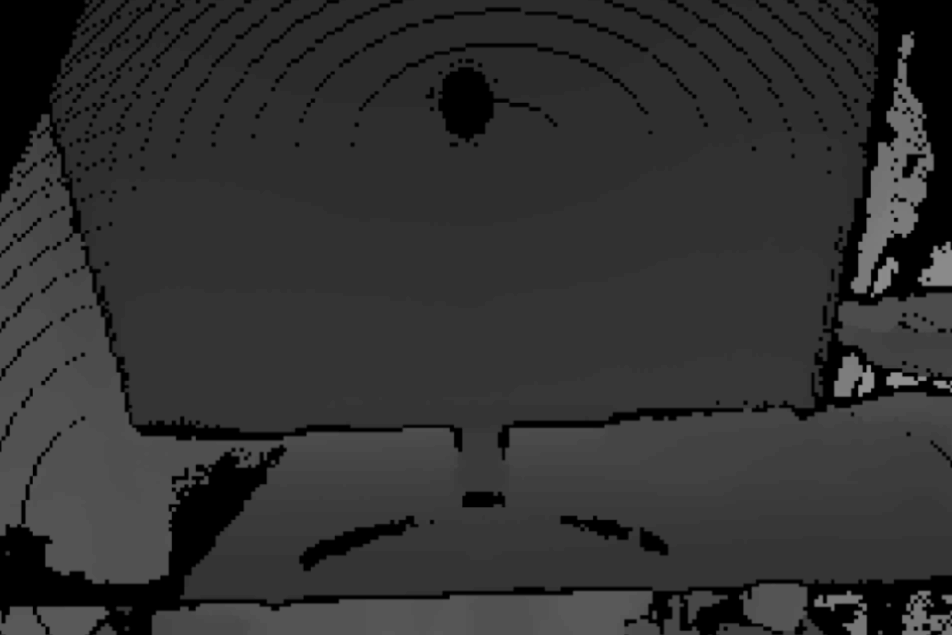}
    \caption{Hololens 2 depth camera.}
    \label{fig:hololens_defense-sub2}
    \end{subfigure}
    \caption{Mitigation via depth sensors on Microsoft Hololens 2. Depth sensors show the screen as flat and lacking details of an image captured of the real location.}
    \label{fig:hololens_defense}
\end{figure}

\paragraph{Additional sensor modalities.} 
The risk of shared state attacks on AR devices can be reduced by leveraging multiple sensor modalities to verify the consistency between the shared state and other accessible sensor data.
For instance, the Microsoft Hololens 2 incorporates Red-Green-Blue (RGB) and depth cameras. 
As shown in Fig.~\ref{fig:hololens_defense}, the depth camera can help identify a computer monitor or photograph, which was key to launching the attacks in Scenarios A and B.
Thus, an automated comparison between the outputs of the depth and RGB cameras can be conducted to detect whether the user is physically present in the actual scene.

To demonstrate this, we designed the following experiment.
We trained a convolutional neural network (CNN) to take color and depth images as input and classify whether the scene is being viewed at the actual physical location or not.
We collected data using the Hololens 2 Sensor Streaming (HL2SS) system~\cite{dibene2022hololens}, which consists of color images saved as 640x480 24-bit RGB images and depth images stored as 640x480 16-bit monochrome images. 
Since the two cameras have different fields of view, we pre-processed the images using the camera's intrinsic parameters to ensure a 1:1 correspondence between pixels in the color and depth images. The data was collected from 15 real scenes, with 300 pairs of color and depth images gathered from each scene.
To simulate an attack, we then collected images taken in front of computer monitors displaying photographs of the same 15 scenes. These monitors varied in size, with diagonals of 11, 32, and 55 inches.

Subsequently, the dataset was partitioned into 12 scenes for training and 3 new unknown scenes for testing (80/20 training/test split).
The CNN was a customized ResNet-18 network~\cite{zhang2023accuracy} modified to incorporate four input channels
(three RGB channels plus depth), followed by a fully connected layer for scene classification of whether the user is physically present in the scene or not.
The results for the multi-model defense are promising, achieving 79.35\%, 79.99\%, and 84.22\% for the F1 score, recall, and precision, respectively.
However, note that not all AR devices, such as iPhones or Android phones, come equipped with depth sensors. This absence of hardware features could potentially restrict the applicability and effectiveness of this defense methodology and further investigation is needed.



\paragraph{Clean slate design.}
Alternatively, the core design of these applications could use traditional security measures to prevent tampering. 
Non-curated shared states (Scenario A and C) could be changed to curated with a permissions system where only trusted users may perform write~\cite{claramunt2023spacemediator}. 
Still, for those applications where crowd-sourcing (non-curation) is desirable, a compromise involving a user reputation system based on past good behavior may prove sufficient, \revision{although this requires oversight by AR providers.}
Additional checks in non-curated applications could be added, such as only accepting appropriately watermarked images with embedded GPS as keys~\cite{lukas2006digital}, 
to prevent false hologram attacks in Scenario C. 

\paragraph{Real space security.}
QR codes printed and placed into the real location can offer a form of locality assurance, particularly if those codes are changed regularly~\cite{roesnermulti}. 
This method ensures that attackers who lack regular physical access to the locations 
will be unable to read holograms remotely. 
Additionally, as we found in \S\ref{sec:geo-evaluation}, read attacks are less successful at greater distances. Thus, we can request that users collect more images at different distances and angles, 
although this places an additional burden on users.


\paragraph{Local moderators.}
AR frameworks with crowd-sourced shared state (Scenario C) may be considered as a form of content hosting (where the content is image keys and hologram values being uploaded). Hence, human moderators may be used to great effect, as in other successful applications like Facebook~\cite{pletikosa2011case}.
However, moderator teams are expensive and must be located close to locations of the uploaded image keys to verify them. 

\section{Related Work}
\label{sec:related}

\paragraph{AR/VR security and privacy overviews.}
Recent overviews~\cite{roesner2021security, GuzmanSecurity} broadly cover existing issues.
Literature covering human factors of multi-user AR also exists~\cite{roesnermulti}, which our work aligns with.
Work on securing AR output in multi-user AR~\cite{ruth2019secure} is orthogonal in that it focuses on content sharing for holograms given their locations, whereas we study how these locations are determined.
The global shared state scenarios also intersect with geospatial information services security covered in~\cite{BertinoGIS}.

\paragraph{AR leakage vectors.}
Prior research~\cite{luo2022holologger, chenguser, romero2023gaitguard} has highlighted the issue of unauthorized acquisition of sensitive information from AR/VR devices. 
Several studies~\cite{meyer2018location, trimananda2021ovrseen} demonstrate the feasibility of inferring the user's location by analyzing network traffic information. 
Other works~\cite{slocum23going, tricomi2023you} establish the ability to deduce keystrokes based on the user's head motions or other user interactions based on performance counters~\cite{zhang2023its}.
Several studies~\cite{gopalhidden, nair2023unique} shows that attackers can exploit sensor-based side-channel leakages to exfiltrate sensitive information.
However, none of these investigate attacks on the shared state in multi-user AR as we do.

\paragraph{Computer vision attacks.}

AR uses computer vision techniques as part of its foundation, and thus such attacks could apply. 
Such work includes software~\cite{wei2020leaky, guo2023white, guo2024artwork, li2023sibling, chen2023dynamic} and hardware~\cite{zhang2021stealing,zhu2023campro, fpga_stealing_cnn} based attacks on machine learning models.
While our work uses photographs, screens, manipulated images, and GPS to trick computer vision systems, attacks using additional hardware like lasers have also been explored~\cite{yan2022rolling}.
SLAM attacks also exist~\cite{SLAMaliasing, SLAMautooverview, nikkhoo2023pimbot} and could impact AR systems.
Our work takes inspiration from these to show computer vision attacks can cascade into interesting behaviors in the AR domain rather than general object detectors or autonomous vehicles.

\paragraph{Sensor spoofing and confusion.}
Our work uses GPS spoofing by simply altering the metadata stored in plain text.
While not necessary for our attacks, more sophisticated GPS spoofing
has reached widespread use~\cite{GPSspoof}. 
Tricking IMU sensors was not done in this work but is possible with acoustic waves~\cite{poltergeist, perturbsensor} and is an interesting future direction.

\paragraph{AR/VR threat mitigation.}

Defenses against user-manipulated input data, such as image manipulation~\cite{ZhouImgManip, ImgManip, liu2022segment,9798870}, have become sophisticated in recent years. 
GPS spoofing mitigation~\cite{jafarnia2012gps} focuses on real-time mitigation, but frameworks like Mapillary provide the ability to upload batched imagery at later times for user convenience, and thus such mitigations may not be directly applicable. 
The most effective mitigation is likely to come in the form of permissions systems like in~\cite{claramunt2023spacemediator}, but these will require non-curated shared states to become curated.

\section{Conclusions}
\label{sec:conclusions}
As AR become ubiquitous, there is growing need for research into security and privacy risks unique to AR, especially multi-user AR.
This paper introduced and explored attacks on multiple shared state AR applications and frameworks.
Specifically, we show that the basic use of GPS and camera images is insufficient to accurately establish the location of an AR device and hence what holograms should be writeable/readable by a user.
We proposed a threat model that applies to several different scenarios and demonstrated them on off-the-shelf AR systems in a variety of environments.
Simple defenses such as image manipulation detection or the use of multi-modal sensors can help, but further investigation of other defenses, such as defining map update policies by trusted users or fraud detection, is needed.



 \section*{Acknowledgments}
We greatly thank the anonymous shepherd and reviewers for their helpful suggestions on the paper.
This work was partially supported by
the NSF grants CNS-1942700, CNS-2053383, CCF-2212426,
and a Meta faculty research award.

{\bibliographystyle{plain}
\small \bibliography{main}}

\appendix
\label{sec:appendix}
\section*{Appendix}

\section{Additional Scenario A results}
\label{sec:scenario_A_results}

\begin{figure*}[t]
    \centering
    \begin{subfigure}[b]{0.15\textwidth}
        \centering
        \includegraphics[width=\textwidth, keepaspectratio]{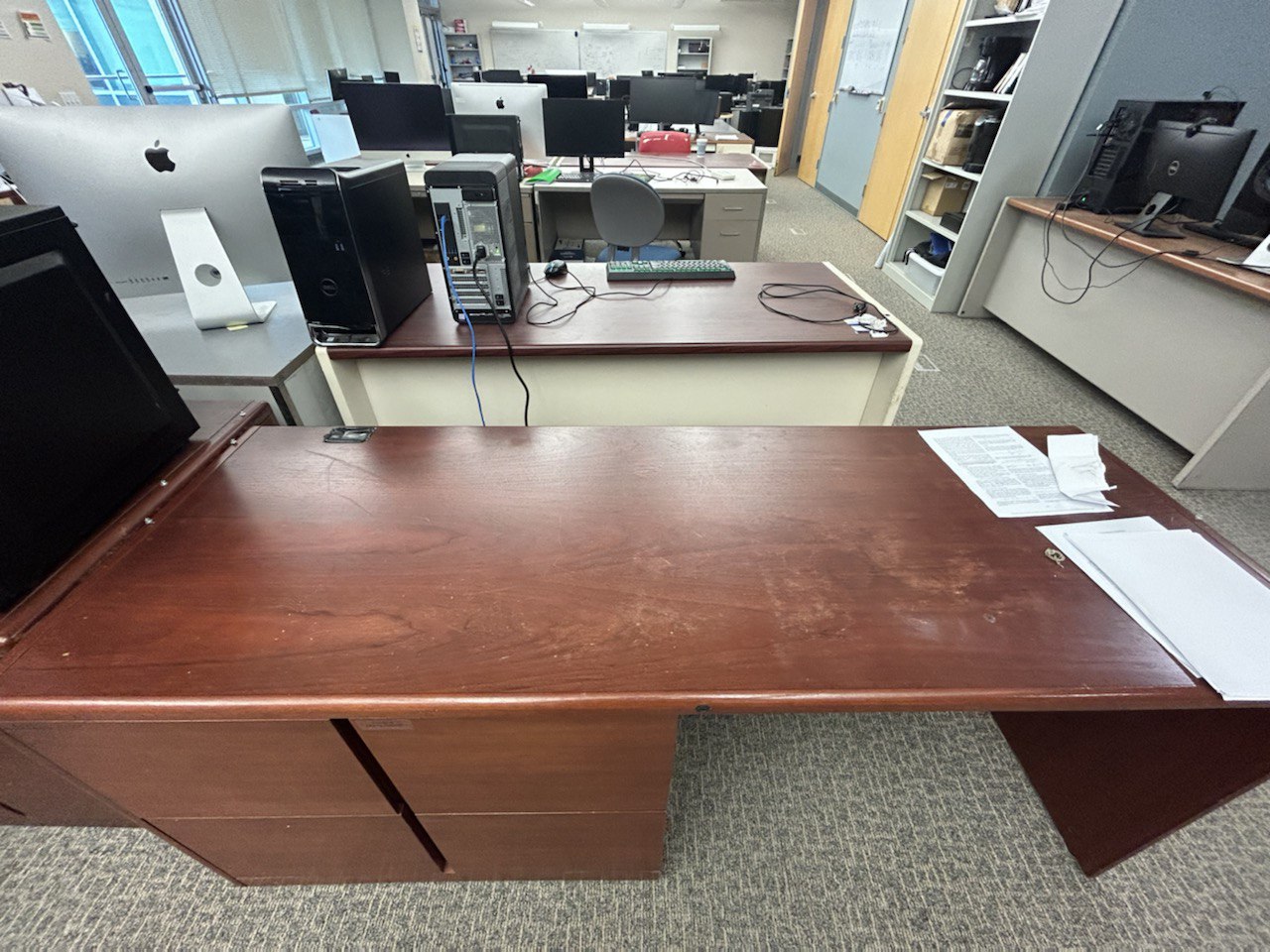}
        \caption{Office desk.}
        \label{fig:indoor_1}
    \end{subfigure}
    \begin{subfigure}[b]{0.15\textwidth}
        \centering
        \includegraphics[width=\textwidth, keepaspectratio]{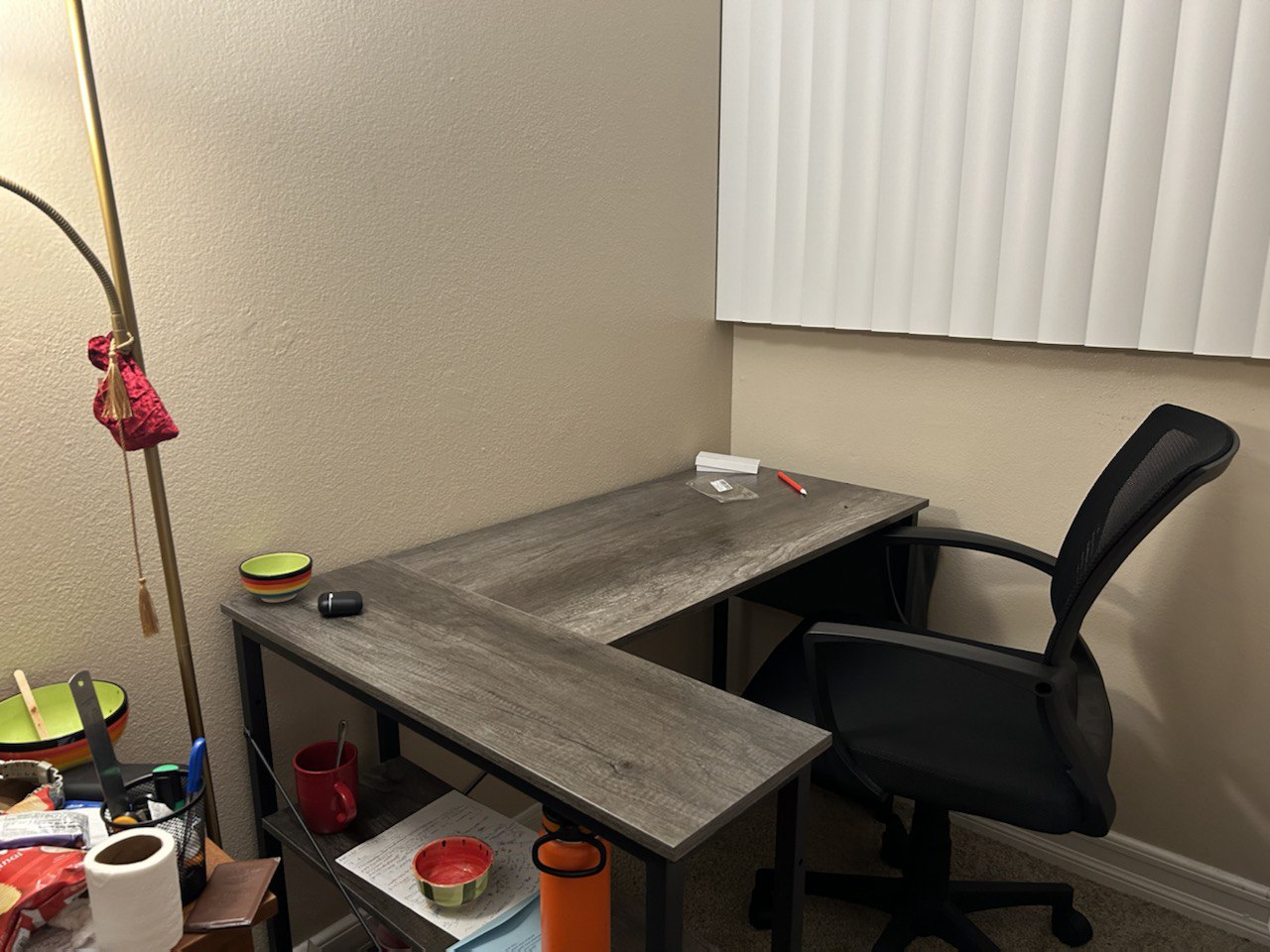}
        \caption{Bedroom desk.}
        \label{fig:indoor_2}
    \end{subfigure}
    \begin{subfigure}[b]{0.15\textwidth}
        \centering
        \includegraphics[width=\textwidth, keepaspectratio]{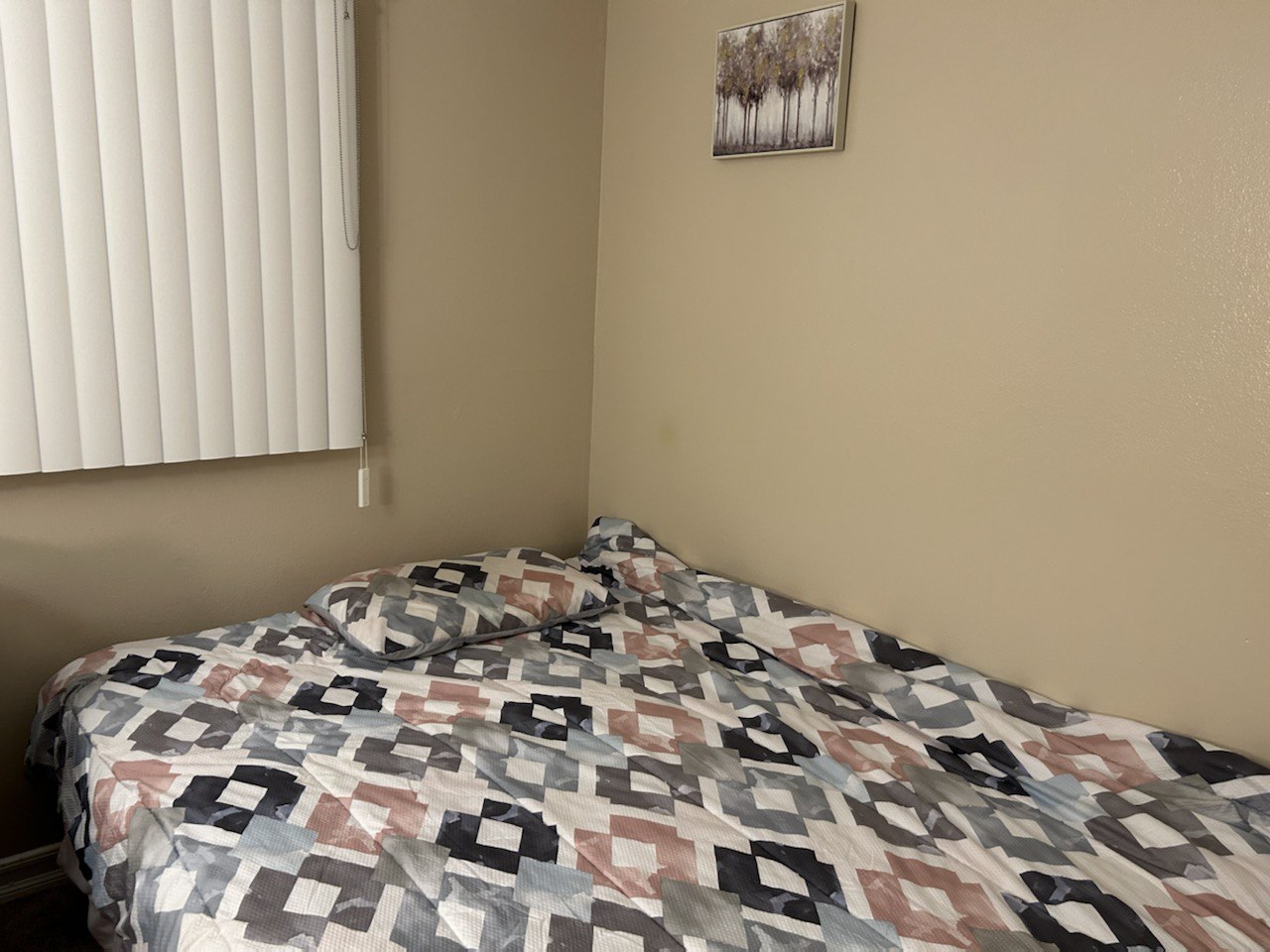}
        \caption{Bedroom bed.}
        \label{fig:indoor_3}
    \end{subfigure}
        \begin{subfigure}[b]{0.15\textwidth}
        \centering
        \includegraphics[width=\textwidth, keepaspectratio]{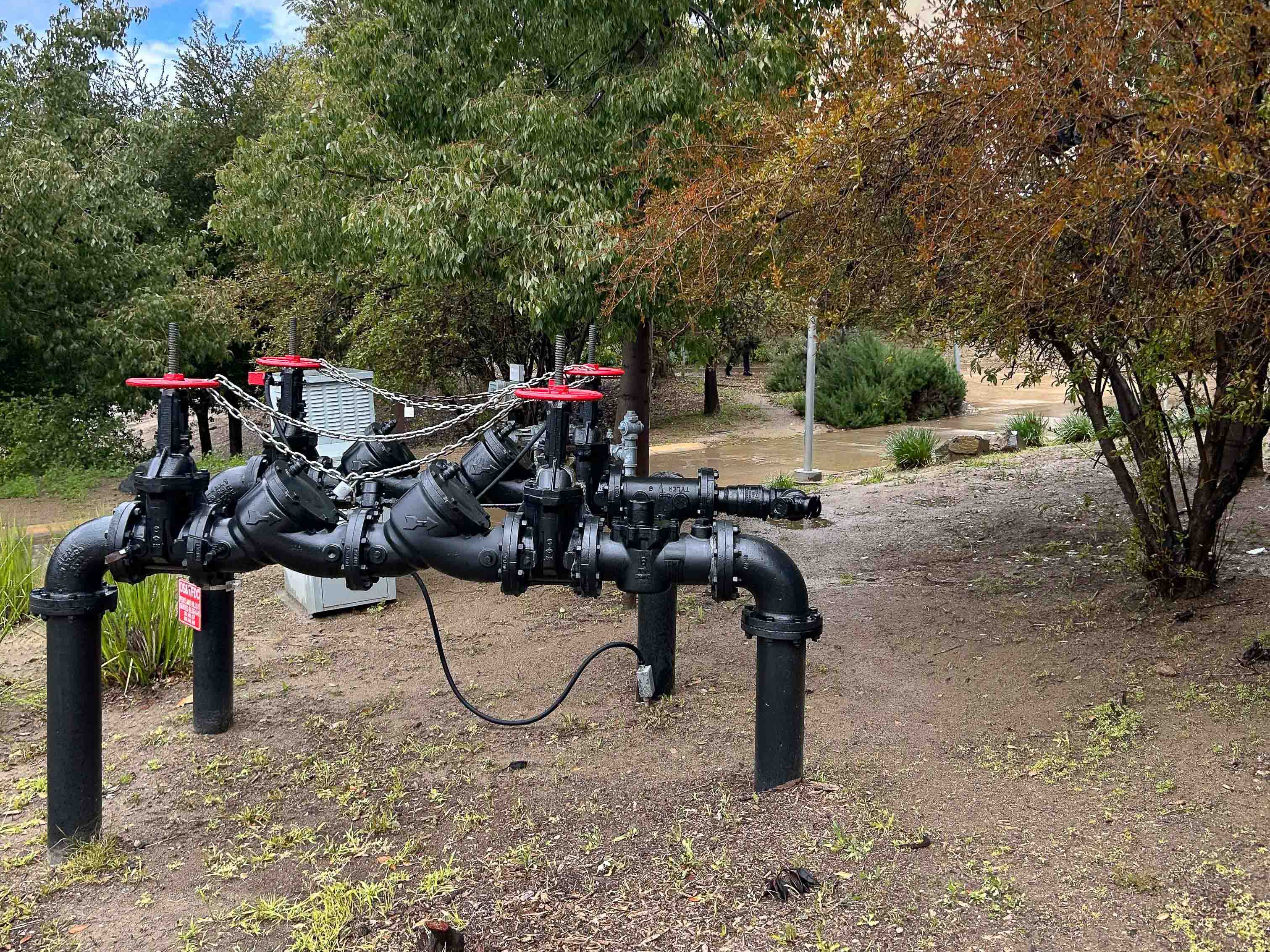}
        \caption{Outdoor garden.}
        \label{fig:garden}
    \end{subfigure}
    \begin{subfigure}[b]{0.15\textwidth}
        \centering
        \includegraphics[width=\textwidth, keepaspectratio]{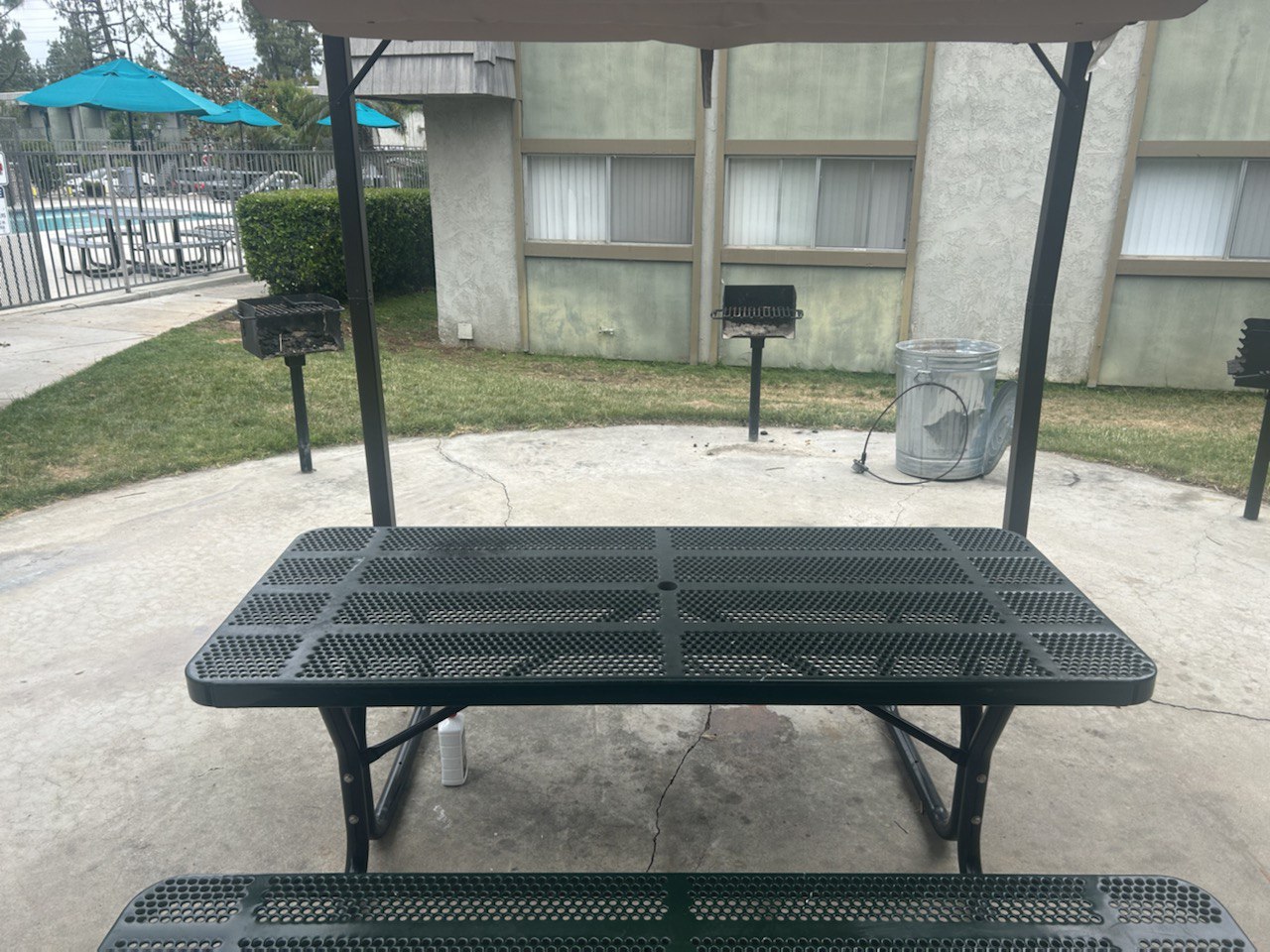}
        \caption{Outdoor BBQ.}
        \label{fig:outdoor_1}
    \end{subfigure}
    \begin{subfigure}[b]{0.15\textwidth}
        \centering
        \includegraphics[width=\textwidth, keepaspectratio]{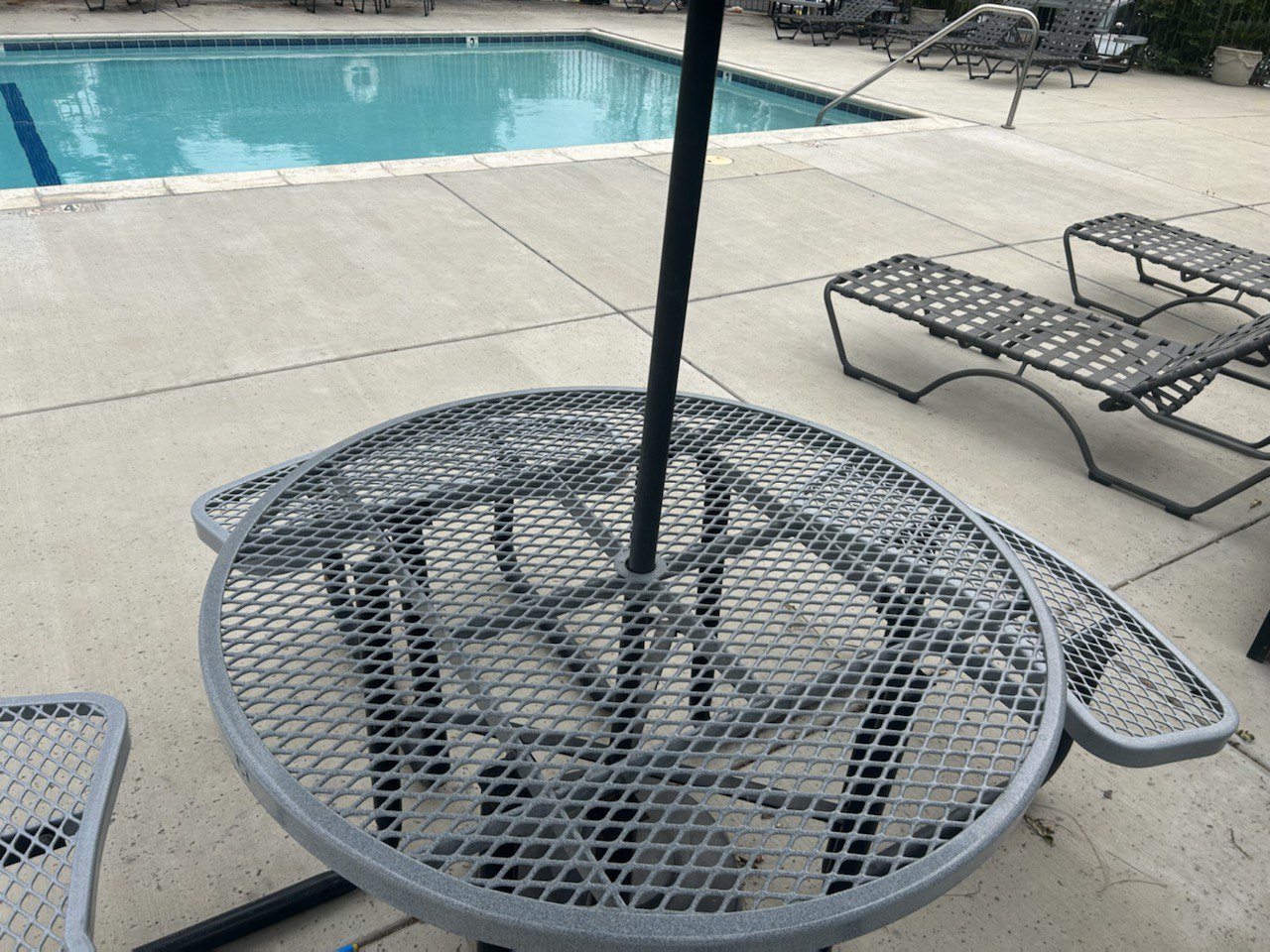}
        \caption{Outdoor pool.}
        \label{fig:pool}
    \end{subfigure}

    \vspace{-0.1in}
    \caption{\revision{Examples of scenes where we conducted our attacks in scenario A.}} 
    \label{fig:arcore_environments}
\end{figure*}

\begin{figure}
  \begin{center}
    \includegraphics[width=0.42\textwidth]{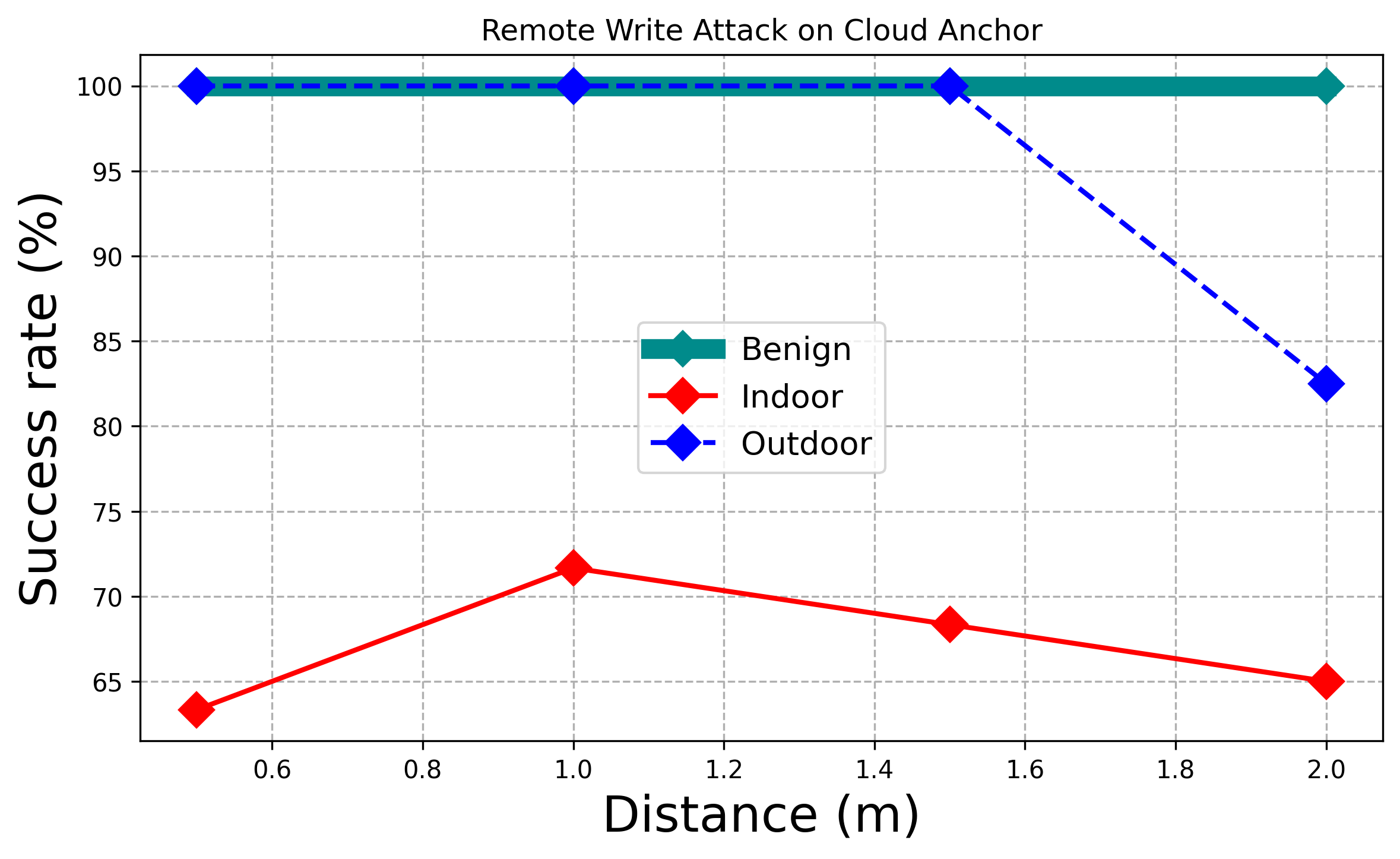}
  \end{center}
  \vspace{-0.2in}
  \caption{\revision{Results of remote write attacks at varying distances for scenario A.}}
  \label{fig:RW_distance_result}
\end{figure}

Fig.~\ref{fig:arcore_environments} shows examples of the environments where we conducted our experiments in Scenario A.
\revision{Fig.~\ref{fig:RW_distance_result} shows the remote write attack as a function of the distance between the AR device's camera and the attacker's image.
It also includes a benign baseline where the attacker writes the hologram while actually being present in the real environment.
Similar to the results in indoor environments (Fig.~\ref{fig:RR_distance_result}), the attack success rate is lower than the benign baseline, as expected, but is still above 60\% in the worst case. }

\section{Details on experimental procedure}
\label{app:experiment_procedure}

\revision{In this section, we provide additional technical details on the experimental procedure for each scenario.
We mention the specific APIs called by the applications, although our attack methodologies apply broadly to AR devices that read and write from shared state.}

\subsection{Scenario A}
\label{app:cloudanchor}

\revision{The hardware and software dependencies for this attack are: an ARCore supported device (\eg Android phone) and access to the ARCore CloudAnchor API.
The attack application was developed using Android Studio version 2022.2.1.}

\revision{
\paragraph{Attacker reads hologram from the remote location.} 
First, a regular user writes a hologram at a physical location using the ARCore CloudAnchor API, for example using the Persistent CloudAnchor demo app~\cite{CloudAnchor_demo}.
This involves opening the app, setting a room code, moving the camera around until the app has scanned the scene sufficiently (following the onscreen prompts), and then tapping on the screen to place the hologram.
Under the hood~\cite{arcore_sdk}, the app sends a \texttt{hostCloudAnchor} request containing an \texttt{Anchor} object to the Firebase server.
The details of the function call are closed source (it calls some underlying NDK C/C++ code), but the documentation indicates that the anchor contains the location and orientation of the hologram and a summary of visual features in the scene near the hologram. 
If the call is successful, the server returns the ID of the hologram (\ie the room code)~\cite{cloud_anchor_intro}.
}

\revision{
After this, an attacker opens the same app and enters the same room code. The attacker points the device's camera at a picture of the previous environment on a screen (\eg a laptop monitor or a TV) and waits for the hologram to appear.
Under the hood, the app sends a \texttt{resolveCloudAnchor} request containing the hologram ID to the Firebase server.
Again, the details of the function call are closed source, but if the call is successful, the server returns the \texttt{Anchor} object corresponding to the ID number and uses this to render the hologram.
To make the attempt successful, the attacker should have a high-quality picture of the environment and also try various distances and slightly different angles when pointing the camera at the picture.
}

\revision{
\paragraph{Attacker writes hologram to the remote location.} 
The attacker opens the app (\eg the CloudAnchor demo app), sets a room code, and points the camera at a picture of the target environment.
In our experiments, we used an Apple monitor to display the picture.
The attacker moves the camera around until the app has scanned the environment sufficiently and taps to place the hologram.
This involves the \texttt{hostCloudAnchor} API call as described previously.
A high-quality picture of the environment is necessary to enhance the success rate of the attack. 
Following the attacker's successful remote write of the hologram, we verified it by having a regular user view the hologram in the physical environment from the picture.
This involves opening the same demo app, entering the room code to call \texttt{resolveCloudAnchor}, and moving the camera around until the hologram appears.
}


\subsection{Scenario B}


\revision{The hardware and software dependencies for this attack are: an ARCore-supported device (\eg Android phone), access to the ARCore Geospatial API, and a GPS emulator~\cite{GPSEmulator}.}

\begin{figure}
    \centering
    \begin{subfigure}[b]{0.15\textwidth}
        \centering
        \includegraphics[width=\textwidth, keepaspectratio]{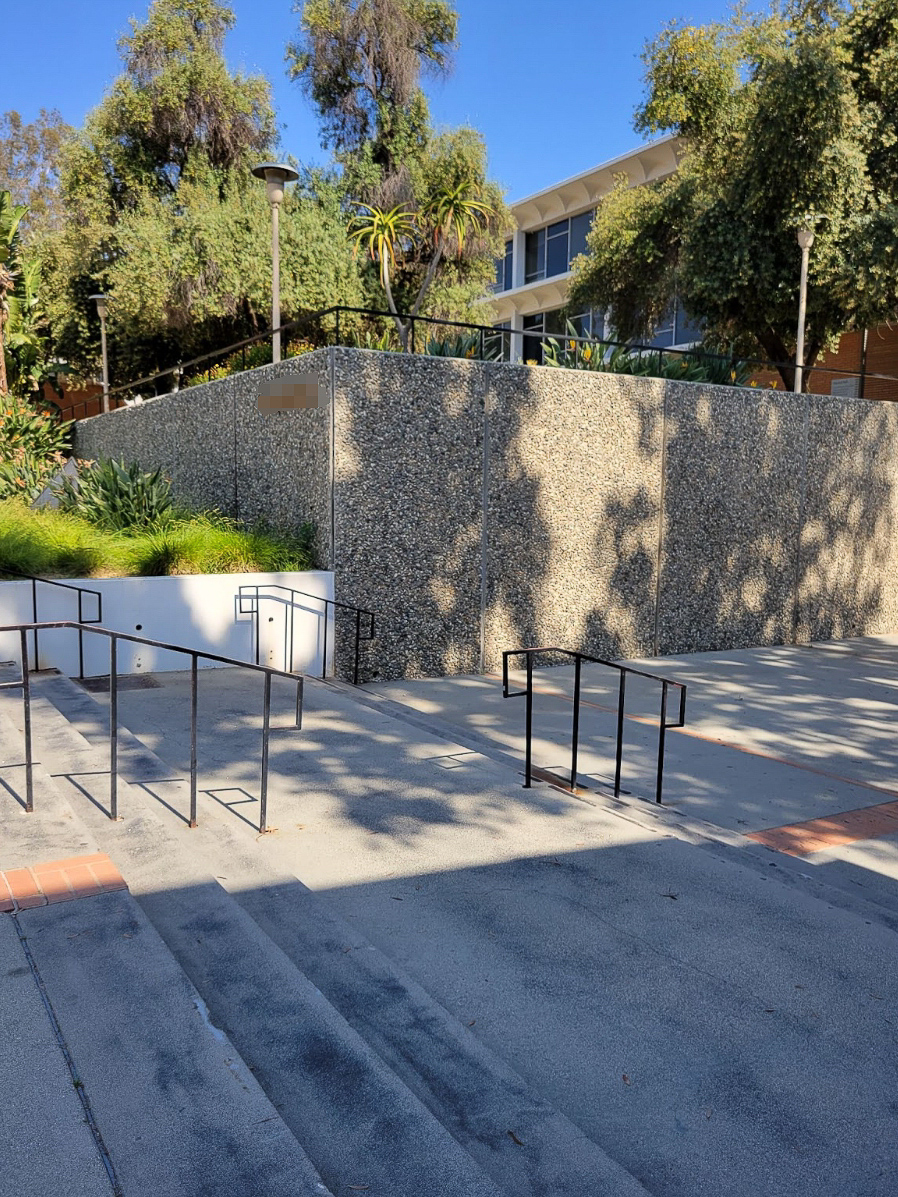}
        \caption{} 
        \label{fig:sub1}
    \end{subfigure}
    \hfill 
    \begin{subfigure}[b]{0.15\textwidth}
        \centering
        \includegraphics[width=\textwidth, keepaspectratio]{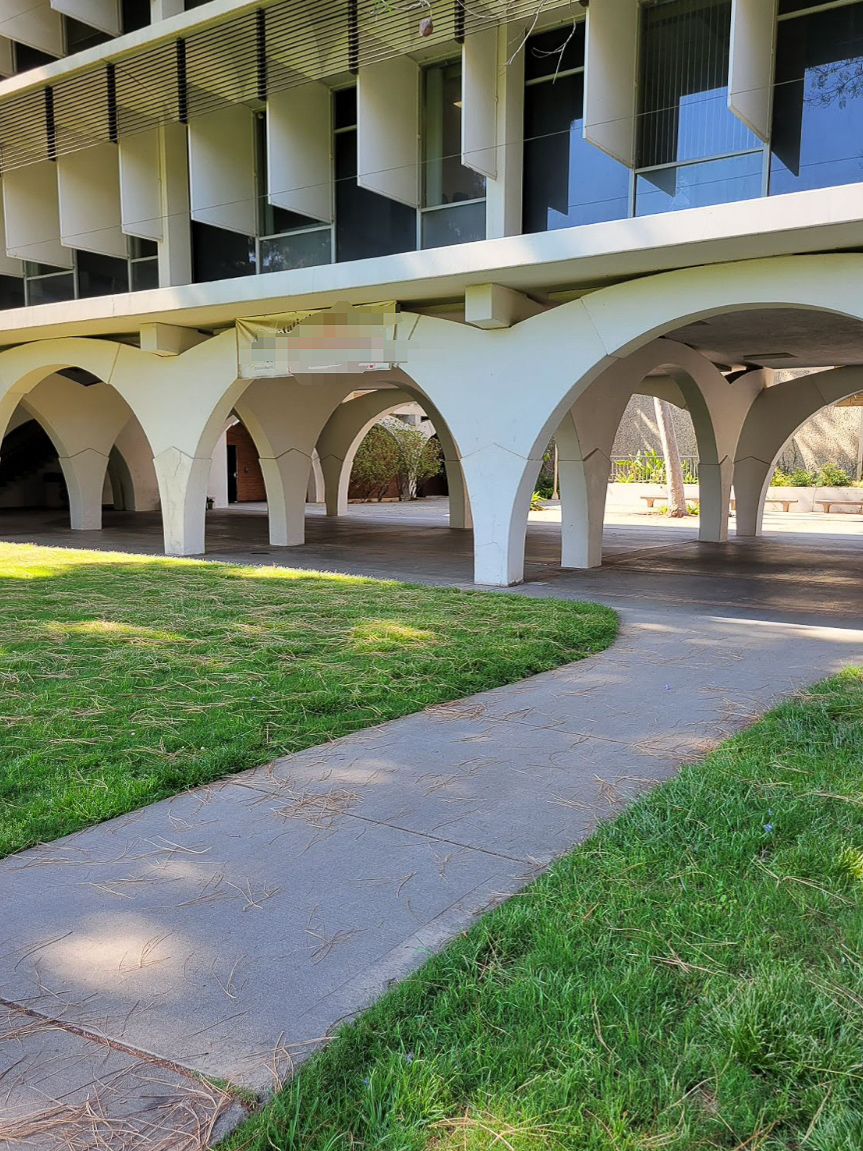}
        \caption{} 
        \label{fig:sub2}
    \end{subfigure}
    \hfill 
    \begin{subfigure}[b]{0.15\textwidth}
        \centering
        \includegraphics[width=\textwidth, keepaspectratio]{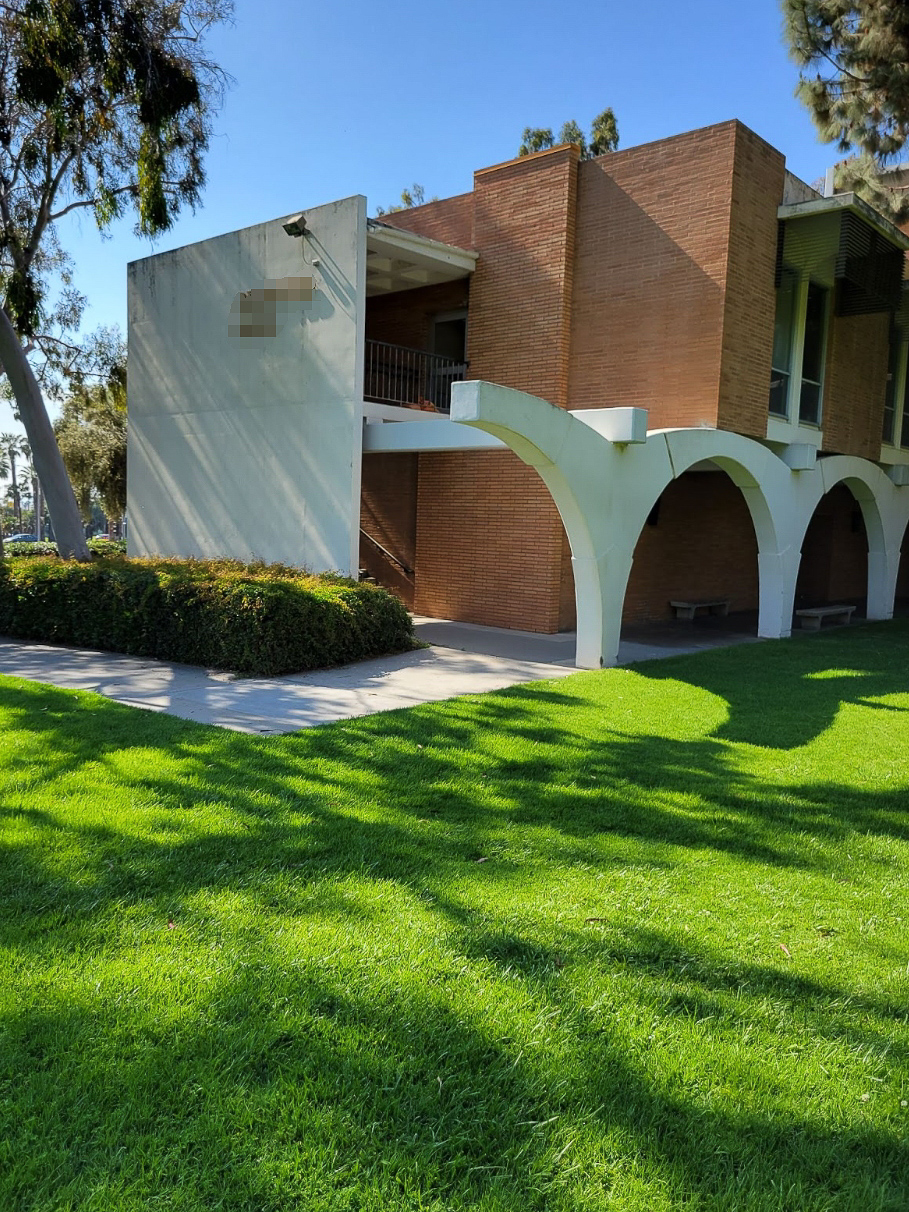}
        \caption{} 
        \label{fig:sub3}
    \end{subfigure}
    \vspace{-0.1in}
    \caption{\revision{Examples of three outdoor scenes where we conducted our attacks in scenario B.}}
    \label{fig:geospatialexample}
\end{figure}

\revision{
\paragraph{(1) Regular user writes hologram to the physical location.}
Before the attacker can launch the remote read attack, a regular user should first write a hologram at a physical location as follows.
The user first aims the device's camera towards the intended outdoor environment. Fig.~\ref{fig:geospatialexample} shows three outdoor environments we evaluated in this work. Once the Geospatial API identifies sufficient features within the scene (this step may take tens of seconds) to localize the device, the user can establish a Geospatial anchor at the precise location by tapping on the screen by following the on-screen prompts~\cite{Geospatial_demo}.
Under the hood~\cite{arcore_sdk}, this involves a call to the \texttt{createAnchorWithGeospatialPose} API function, which saves the latitude, longitude, altitude, and orientation of the hologram (all automatically determined by the location of the user's tap) into an \texttt{Anchor} object.
Subsequently, the GPS coordinates in the \texttt{Anchor} can be recorded and preserved in the shared state.
}

\revision{
\paragraph{(2) Attacker reads hologram from the remote location.}
The attacker first employs a GPS emulator~\cite{GPSEmulator} to fabricate GPS locations on the mobile device corresponding to the locations in step (1) above.
At the same time, the attacker utilizes a monitor to display an image of the desired location from step (1) and directs the mobile device's camera towards it.
To ensure optimal focus, the attacker may need to adjust the device's position, moving forward or backward as necessary.
(In our attack, we find that when the distance between the monitor and the phone is 50 centimeters, our attack succeeds 100\% of the time.)
Together, the GPS and the camera frames help mislead the shared state into believing the attacker is physically present at the target location.
This is done by the Geospatial API's \texttt{Earth} object, which makes calls to the shared state to determine and track the device's location on Earth.
After the device localization is successful (\ie the \texttt{Earth}'s tracking state is not null),
the hologram corresponding to the target location will be rendered on the monitor in the device's display, effectively deceiving the shared state into supplying holograms to be read remotely.
}

\subsection{Scenario C}

\revision{
\paragraph{(1) Image capture.} The attacker uses a smartphone (iPhone 12 at 2532 × 1170 pixels in our experiments) to capture photos of a location that is desired to appear somewhere else.
These photos are automatically geo-tagged with latitude, longitude, time and elevation by the mobile device's operating system.
The user then transfers the photos from the mobile device onto a computer with the ability to run Mapillary's desktop client~\cite{mapillary-uploader} as well as simple scripts written by the attacker. 
}

\revision{
\paragraph{(2) GPS spoofing.} There are two convenient ways to determine the desired GPS data to spoof. The simplest for the attacker is to physically go to the target location that she wishes to write the fake data to, and take real images in a manner similar to the image capture step above.
Then, the attacker can swap the EXIF metadata between the two image sets (from the image capture step and from the GPS spoofing step) to perform the spoof. This can done using a Python script or through manual edits to the image metadata.
For the second set of images, It is important to take the same number of images while moving or walking in the same direction as the first set.
This helps the GPS coordinates match up between the two sets of images.
}

\revision{The second method to determine the desired GPS data is to overwrite the EXIF image metadata manually, using a program like Windows Photo Viewer or a custom script. This is tedious as it requires the attacker to guess the change in GPS coordinates for each image in the set.
}

\revision{
\paragraph{(3) Upload.}
Finally, the attacker uploads the altered image set to the shared state servers using a desktop client (\eg \cite{mapillary-uploader}).
Under the hood~\cite{mapillary-uploader-code},
This follows a standard upload procedure including the image file, metadata, and account information. 
In our experiments, we did this using special accounts that uploaded data to a private sandbox, thus avoiding any impact on regular public users.
Care must be taken that no other EXIF data was removed during any of the previous image capture or GPS spoofing steps, as otherwise the upload may fail.
For example, we found that during the image transfer process between mobile devices and desktops, the timestamps did not transfer and had to be re-added manually. Each step in this process also keeps the images as PNG to avoid lossy compression.
}





\end{document}